\documentclass[11pt,preprintnumbers,
floatfix,superscriptaddress,nofootinbib,showpacs]{revtex4-1}
\pdfoutput=1
\usepackage{amsmath,amssymb,amsthm,amsxtra,overpic,bbm,bm,epsfig,subfigure}
\usepackage{color}
\usepackage{graphicx}
\usepackage{dcolumn}
\usepackage{bm}
\usepackage{epsfig,cancel}
\usepackage{amsmath}
\usepackage{amssymb}
\usepackage{graphicx,bm}
\usepackage{epstopdf}
\usepackage{color}
\usepackage{url}
\usepackage{breakurl}
\usepackage{collref}

\allowdisplaybreaks[4]
\numberwithin{equation}{section}

\def\beq{\begin{equation}}
\def\eeq{\end{equation}}
\def\bsp#1\esp{\begin{split}#1\end{split}}
\newcommand{\be}{\begin{equation}}
\newcommand{\ee}{\end{equation}}
\newcommand{\bea}{\begin{eqnarray}}
\newcommand{\eea}{\end{eqnarray}}

\newbox\charbox
\newbox\slabox
\newbox\ourfigbox
\def\s#1{{      % Feynman slash
        \setbox\charbox=\hbox{$#1$}
        \setbox\slabox=\hbox{$/$}
        \dimen\charbox=\ht\slabox
        \advance\dimen\charbox by -\dp\slabox
        \advance\dimen\charbox by -\ht\charbox
        \advance\dimen\charbox by \dp\charbox
        \divide\dimen\charbox by 2
        \raise-\dimen\charbox\hbox to \wd\charbox{\hss/\hss}
        \llap{$#1$}
}}

\def\ksl{\not{\hbox{\kern-2.3pt $k$}}}

\def\Ord{{\cal O}}

\def\spa#1.#2{\left\langle#1\,#2\right\rangle}
\def\spb#1.#2{\left[#1\,#2\right]}
\def\lor#1.#2{\left(#1\,#2\right)}
\def\sand#1.#2.#3{%
\left\langle\smash{#1}{\vphantom1}^{-}\right|{#2}%
\left|\smash{#3}{\vphantom1}^{-}\right\rangle}

\def\n{n}    % light cone vector

\def\n3lo{N$^3$LO}

\newfont{\scyr}{wncyr10 scaled 550}

\newcommand{\red}{\textcolor{black}}
\newcommand{\nbrk}{\nonumber\\}

\def\sss{\scriptscriptstyle}

\begin{document}
\preprint{MIT-CTP-4757}

\title{Diphoton excess at 750 GeV: $gg$ fusion or $q\bar{q}$ annihilation?}
\author{Jun Gao}
\affiliation{High Energy Physics Division, Argonne National Laboratory, Argonne, Illinois 60439, USA}
\email{jgao@anl.gov}
\author{Hao Zhang}
\affiliation{Department of Physics, University of California, Santa Barbara, California 93106, USA}
\email{zhanghao@physics.ucsb.edu}
\author{Hua Xing Zhu}
\affiliation{Center for Theoretical Physics, Massachusetts
  Institute of Technology,
Cambridge, MA 02139, USA}
\email{zhuhx@mit.edu}

\begin{abstract}
\noindent Recently, ATLAS and CMS collaboration reported an excess in the
diphoton events, which can be explained by a new resonance with mass
around 750 GeV. In this \red{work,} we explored the possibility of
identifying if the hypothetical new resonance is produced through gluon-gluon
fusion or quark-antiquark annihilation, or tagging the beam. 
Three different observables for beam tagging, namely the rapidity and
transverse momentum distribution of the diphoton, and one tagged
bottom-jet cross section{, are proposed}. \red{C}ombining the information gained
from these observables, a clear distinction of the production
mechanism for the diphoton resonance is promising.
\end{abstract}
\maketitle

\section{Introduction}
\label{sec:introduction}

Very recently\red{, both ATLAS and CMS collaboration
present their new results at the LHC Run 2.} Although most
of the measurements can still be fit in the Standard Model (SM)
framework nicely, some intriguing excesses \red{are} reported. Of
particular interest is the diphoton excess around 750 GeV seen by both
collaborations. 
\red{The ATLAS collaboration reports an excess above the 
standard model (SM) diphoton background with a local
(global) significance of 3.9 (2.3) $\sigma$ \cite{ATLAS-CONF-2015-081}. The CMS
collaboration, with a little less integrated luminosity, also 
reports an excess at 760 GeV with a local (global) 
significance of 2.6 (a little less than 1.2) $\sigma$ \cite{CMS-PAS-EXO-15-004}.}

Though further data is required in order to establish the existence
of a new resonance or other beyond the \red{SM} (BSM)
mechanism responsible for the diphoton excess, significant theoretical
efforts have been put into explaining the possible diphoton excess in
various BSM scenarios \cite{Harigaya:2015ezk,*Mambrini:2015wyu,*Backovic:2015fnp,*Angelescu:2015uiz,*Nakai:2015ptz,*Knapen:2015dap,*Buttazzo:2015txu,*Pilaftsis:2015ycr,*Franceschini:2015kwy,*DiChiara:2015vdm,*McDermott:2015sck,*Ellis:2015oso,*Low:2015qep,*Bellazzini:2015nxw,*Gupta:2015zzs,*Petersson:2015mkr,*Molinaro:2015cwg,*Costa:2015llh,*Dutta:2015wqh,*Cao:2015pto,*Yamatsu:2015oit,*Matsuzaki:2015che,*Kobakhidze:2015ldh,*Martinez:2015kmn,*Cox:2015ckc,*Becirevic:2015fmu,*No:2015bsn,*Demidov:2015zqn,*Gopalakrishna:2015dkt,*Chao:2015ttq,*Fichet:2015vvy,*Fichet:2015vvy,*Bian:2015kjt,*Chakrabortty:2015hff,*Ahmed:2015uqt,*Agrawal:2015dbf,*Csaki:2015vek,*Falkowski:2015swt,*Aloni:2015mxa,*Bai:2015nbs,*Gabrielli:2015dhk,*Benbrik:2015fyz,*Kim:2015ron,*Alves:2015jgx,*Megias:2015ory,*Carpenter:2015ucu,*Chao:2015nsm,*Arun:2015ubr,*Han:2015cty,*Chang:2015bzc,*Chakraborty:2015jvs,*Ding:2015rxx,*Han:2015dlp,*Han:2015qqj,*Luo:2015yio,*Chang:2015sdy,*Bardhan:2015hcr,*Feng:2015wil,*Wang:2015kuj,*Cao:2015twy,*Huang:2015evq,*Liao:2015tow,*Heckman:2015kqk,*Dhuria:2015ufo,*Bi:2015uqd,*Kim:2015ksf,*Berthier:2015vbb,*Cho:2015nxy,*Cline:2015msi,*Chala:2015cev,*Barducci:2015gtd,*Boucenna:2015pav,*Murphy:2015kag,*Hernandez:2015ywg,*Dey:2015bur,*Pelaggi:2015knk,*deBlas:2015hlv,*Belyaev:2015hgo,*Dev:2015isx,*Huang:2015rkj,*Moretti:2015pbj,*Patel:2015ulo,*Badziak:2015zez,*Chakraborty:2015gyj,*Cao:2015xjz,*Altmannshofer:2015xfo,*Cvetic:2015vit,*Gu:2015lxj,*Allanach:2015ixl,*Davoudiasl:2015cuo,*Craig:2015lra,*Das:2015enc,*Cheung:2015cug,*Liu:2015yec,*Zhang:2015uuo,*Casas:2015blx,*Hall:2015xds}. While the models proposed vary
significantly, there are some common features shared by most of
them.
Due to the quantum number of
photon pair, most of the proposals suggest that the excess is either due to
gluon-gluon fusion or quark-antiquark annihilation. Different production
mechanisms can lead to very different UV models. Knowing the actual
production mechanism responsible for the potential excess is of great
importance for understanding the underlying theory. Unfortunately, very little
can be said from the current data, except some considerations based on
the consistence of experimental data from \red{the LHC Run 1 and Run 2}.

In this work, we shall study the following problem: if the diphoton excess
persists in future data, and the existence of a new resonance is
established, is it possible to distinguish  different production
mechanisms with enough amount of data? One can compare this question with the
more frequently asked question, namely, how to tell whether an
energetic  hadronic jet in the final state is due to a quark or a
gluon produced from hard scattering. This is also known as the quark
and gluon jet tagging problem, see e.g. refs.~\cite{1106.3076,1408.3122,1501.04794,1512.05265}~\footnote{Somewhat
  related discussion have also been made in the literatures about the
  color content of BSM resonance production~\cite{0908.3688,1108.2396}, and the tagging of initial-state
  radiation~\cite{1101.0810}.}. One can view the
question of differentiating the $gg$ fusion and $q\bar{q}$
annihilation mechanism as a final-state-to-initial-state crossing of
the quark and gluon jet tagging problem. For this reason we will call
it the quark and gluon beam tagging problem in this work, or beam
tagging for short. While our current work in the beam tagging problem was
motivated by the diphoton excess, we believe that our results
will be useful even if the excess disappear after more data is
accumulated, because a bump might eventually show up at a different
place or/and in a different channel.

An important feature of the beam tagging problem is that most of the
QCD radiations from the initial-state partons are in the forward
direction, and therefore are hard to make used of. This is contrasted
with final-state jet tagging, in which the information of QCD
radiations in the jet play crucial role in identifying the partonic
origin of the jet. This feature makes the beam tagging problem
difficult. Based on the consideration of general properties of initial-state QCD radiations, we explore different observables which are useful for  the beam
tagging problem. Firstly, we consider the rapidity distribution of the
diphoton system. It is well-known from Drell-Yan production that for
$q\bar{q}$ initial state, contribution from valence quark and sea
quark can have different shape in rapidity distribution.  Using this
information, we find that it is possible to distinguish the valence
quark scattering from sea quark or gluon scattering. Secondly, we
consider the transverse momentum~($Q_T$) distribution of the diphoton
system. It is well-known that $Q_T$ distribution of a color neutral
system exhibits a Sudakov peak at low $Q_T$ due to initial-state QCD
radiation. Interestingly, the strength of initial-state radiation
differs for quark or gluon induced hard scattering and leads to
substantial difference in the position of the Sudakov peak. Using this
information, it is possible to distinguish light quark scattering from
 bottom quark or gluon scattering. Lastly, to further differentiate
 bottom quark induced or gluon induced scattering, we consider tagging
 a $b$-quark jet in the final state.

\section{Three methods for the beam tagging problem}
\label{sec:750-gev-diphoton}

We consider the following effective operators with an additional
singlet scalar $S$,
\begin{align}
  \label{eq:1}
  {\cal L}_{\rm eff}=- \frac{1}{4} \frac{\alpha_s}{3 \pi \Lambda_g} S 
  G^a_{\mu\nu}G^{a,\mu\nu} +\sum_{f=u,s,d,c,b}\Big( \frac{v}{\Lambda_f}
S  \bar{q}_f \red{q}_f + \rm{h.c.} \Big).
\end{align}
There could also be effective operators with a pseudo scalar. But their
long distance behavior is in-distinguishable from the scalar
case. Also the scalar has to couple to photon in order to be able to
decay to diphoton. But that is irrelevant to \red{most of} our discussion.

Thanks to QCD factorization, the hadronic production cross section for
$S$ can be written as
\begin{align}
  \label{eq:2}
  \sigma^{(i)}_0 =\tau  \int^1_\tau \frac{dx}{x}\, \Big( f_{i/N_1}(\tau/x)
  f_{\bar{i}/N_2}(x) + (i \leftrightarrow \bar{i})\Big)\hat{\sigma}^{(i)}_0\red{,}
\end{align}
where $\tau = M^2_S/E^2_{\rm CM}$.
The operator in Eq. (\ref{eq:1}) leads to the following partonic cross
section to the scalar production: 
\begin{align}
  \label{eq:3}
  {\hat\sigma}^{(g)}_0  = &\,\frac{\pi}{16(N_c^2 -1)} \Big(\frac{\alpha_s}{3\pi \Lambda_g}\Big)^2\red{,}
\nbrk
  {\hat\sigma}^{(q)}_0  = &\,\frac{\pi}{2 N_c} \Big(
                            \frac{v}{\Lambda_q M_S}\Big)^2\red{.}
\end{align}

\subsection{Rapidity distribution}

\begin{figure}[h]
  \begin{center}
  \includegraphics[width=0.5\textwidth]{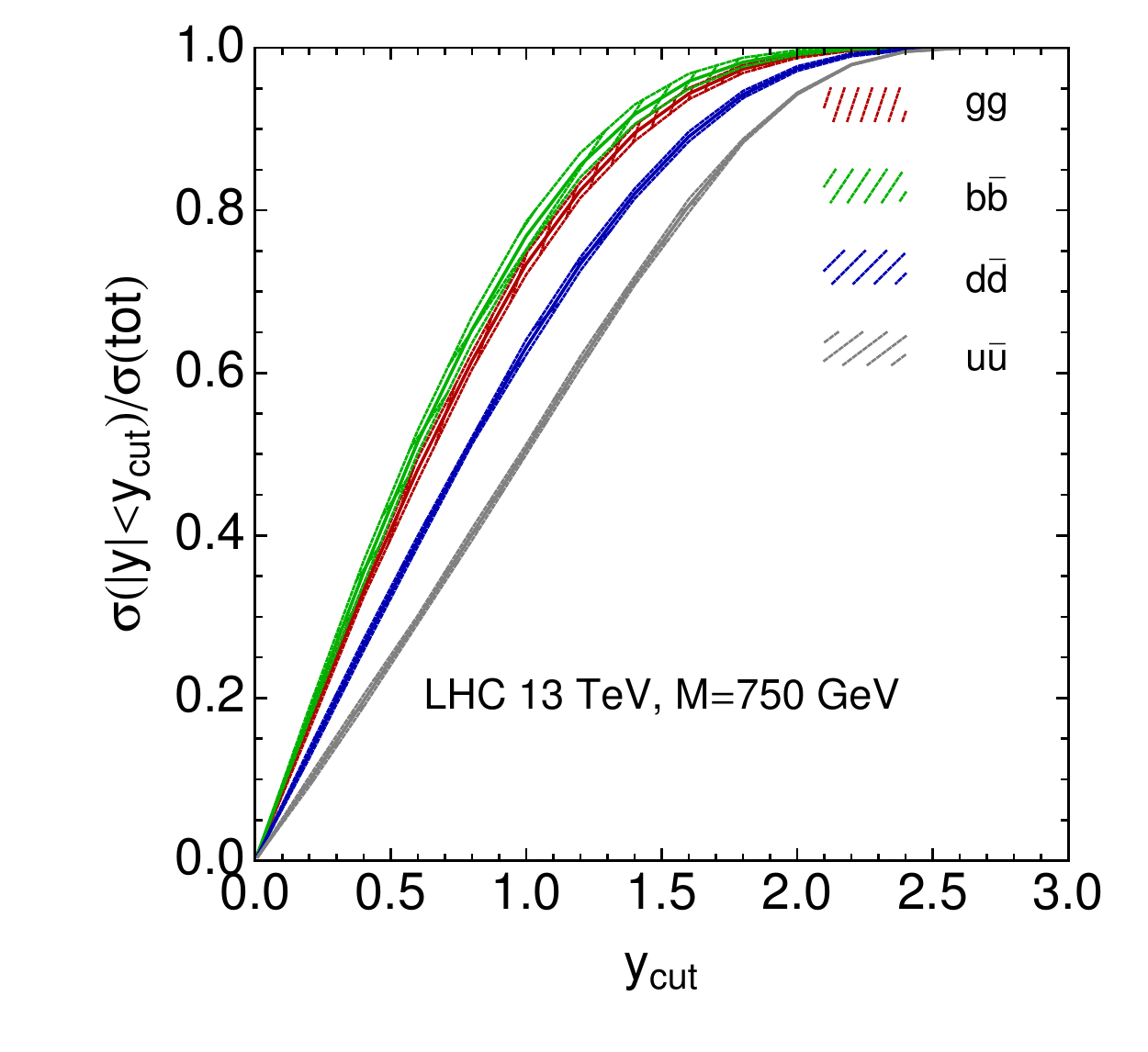}
  \end{center}
  \vspace{-2ex}
  \caption{\label{fig:y}
   Centrality ratio, defined as ratio of cross section in central
   rapidity region $|y|<y_{cut}$ and the total cross section, as a
   function of $y_{cut}$.}
\end{figure}

It is well-known that for $W$ and $Z$ boson production in the
\red{SM}, contributions from different partonic channels
have different shapes in rapidity distribution of the boson. Valence-quark
contributions have a double shoulder structure while the
sea-quark contributions peak in the central region due to
different slopes of the \red{parton distribution functions (PDFs)}
with respect to Bjorken $x$. The
results are similar for a resonance of $750$ GeV produced
at \red{13 TeV LHC}. One way to quantify the shape of rapidity
distribution is to use the centrality ratio, which is defined
as ratio of cross sections in central rapidity region $|y|<y_{cut}$
and the total cross sections. In FIG. \ref{fig:y} we show
the centrality ratio as a function of $y_{cut}$ for a $750$ GeV
resonance produced through different parton combinations at leading order (LO).
The hatched bands show the corresponding 68\% \red{confidence level (C.L.)} PDF uncertainties
as calculated according to the PDF4LHC recommendation \cite{Butterworth:2015oua}, which
are small especially for the valence-quark contributions. The ratios approach
one when $y_{cut}$ approaching endpoint of the rapidity distribution
$\sim 2.8$. As expected the valence-quark contributions have
smaller values for the ratio than ones from gluon or bottom quarks.
The ratios are very close for gluon and bottom-quark or other sea-quark
contributions, since the sea-quark PDFs are mostly driven by the gluon
through DGLAP evolution. Taken $y_{cut}$ to be 1, the centrality ratios
are 0.74, 0.77, 0.63 and 0.50, for $gg$, $b\bar b$, $d \bar d$ and $u\bar u$
channels respectively. Assuming most of the experimental systematics will cancel
in the ratio and with high statistics, it will be possible to discriminate underlying
theory with production initiated by valence quarks and by gluon or sea quarks.
Higher-order perturbative corrections may change above numbers which depend on the
full theory.       

\subsection{Diphoton at small transverse momentum}
\label{sec:diphoton-at-small}

We next consider the transverse momentum $Q_T$ of the diphoton
system. In the SM, transverse momentum resummation for diphoton has
been considered at Next-to-Next-to-Leading Logarithm~(NNLL) level~\cite{1505.03162}. Fully differential
distribution is also known at fixed next-to-next-to-leading order (NNLO)~\cite{1110.2375}. Here we
consider the case where the diphoton originates from the decay of a
new resonance at 750 GeV. At LO in QCD, $Q_T$ is exactly zero due to momentum conservation
in the transverse plane. However, as is well-known from study of
Drell-Yan lepton pair transverse-momentum distribution, $Q_T$ is not
peaked at zero 
but rather at finite transverse momentum. The shift from $Q_T=0$ to
non-zero value is mostly due to initial-state QCD
radiation. For example, if the diphoton is produced from $gg$ fusion,
the initial-state gluon in one proton can split into two gluons before
colliding with the gluon from the other proton. The diphoton system is
pushed to non-zero $Q_T$ as a result of the splitting
process. For large $Q_T$, the strong coupling is small and
perturbative expansion works well. However, when $Q_T$ is much smaller
than $M_S$, large logarithms of the ratio between $M_S$ and
$Q_T$ could arise, which spoils the convergence of the perturbative
series. As an example, at NLO, the partonic cross section  for the
$Q_T$ distribution of the diphoton system at leading power in $Q_T^2/M^2_S$ can be written as
\begin{align}
  \label{eq:4}
\left.\frac{d\sigma^{(g)}}{dQ^2_T}\right|_{Q^2_T>0} =\left(\frac{\alpha_s}{4\pi} \right) &\,2  \tau \hat{\sigma}^{(g)}_0
                                   \sum_{a,b} \int^1_0 dx_1 \int^1_0
                                   dx_2 \, \delta( x_1 x_2 - \tau)
                                   \int^1_{x_1} \frac{d\xi_1}{\xi_1}
                                   f_{a/N_1}(x_1/\xi_1)  \int^1_{x_2}
                                   \frac{d\xi_2}{\xi_2}
                                   f_{b/N_2}(x_2/\xi_2)
\nbrk
&\, \times \Bigg\{ \delta_{ag}\delta_{bg} \delta( 1 - \xi_1)
  \delta( 1 - \xi_2)  \Bigg[4C_A
  \frac{\ln(M^2_S/Q^2_T)}{Q^2_T} -\Bigg( \frac{22}{3} C_A -
  \frac{4}{3} N_f\Bigg) \frac{1}{Q^2_T} \Bigg]
\nbrk
&\, + \frac{2 P_{ga}(\xi_1) }{Q^2_T} \delta_{bg} \delta(1 - \xi_2)  +
  \frac{  2 P_{gb}(\xi_2) }{Q^2_T} \delta_{ag} \delta(1 - \xi_1)
  \Bigg\} + \Ord(\alpha_s^2)
\end{align}
for $gg$-fusion production. Similarly, for $q\bar{q}$ induced diphoton production, we have
\begin{align}
  \label{eq:9}
  \left.\frac{d\sigma^{(q)}}{dQ^2_T}\right|_{Q^2_T>0} = &\,
\left(\frac{\alpha_s}{4\pi} \right) \tau \hat{\sigma}^{(q)}_0
                                   \sum_{a,b} \int^1_0 dx_1 \int^1_0
                                   dx_2 \, \delta( x_1 x_2 - \tau)
                                   \int^1_{x_1} \frac{d\xi_1}{\xi_1}
                                   f_{a/N_1}(x_1/\xi_1)  \int^1_{x_2}
                                   \frac{d\xi_2}{\xi_2}
                                   f_{b/N_2}(x_2/\xi_2)
\nbrk
&\, \times \Bigg\{ \delta_{aq}\delta_{b\bar{q}} \delta( 1 - \xi_1)
  \delta( 1 - \xi_2)  \Bigg[4C_F
  \frac{\ln(M^2_S/Q^2_T)}{Q^2_T} - 6 C_F\frac{1}{Q^2_T} \Bigg]
\nbrk
&\, + \frac{ 2P_{qa}(\xi_1) }{Q^2_T} \delta_{\bar{q}g} \delta(1 - \xi_2)  +
  \frac{ 2P_{gb}(\xi_2) }{Q^2_T} \delta_{ag} \delta(1 - \xi_1)  \Bigg\}
  + \Big( q \leftrightarrow \bar{q} \Big)   + \Ord(\alpha_s^2)\red{,}
\end{align}
where $P_{ij}(z)$ are the LO QCD splitting functions:
\begin{align}
  \label{eq:10}
  P_{qq}(z) = &\, C_F\Big[ \frac{1+z^2}{1 - z} \Big]_+\red{,} 
\nbrk
P_{qg}(z)  =&\, \frac{1}{2} \big[ (1-z)^2 + z^2 \big]\red{,}
\nbrk
P_{gg}(z) = &\, 2 C_A \frac{(1 -z + z^2)^2}{z} \Big[ \frac{1}{1-z}
              \Big]_+ + \Big( \frac{11}{6} C_A - \frac{1}{3} N_f \Big)\red{,}
              \delta( 1 -z)
\nbrk
P_{gq}(z) = &\,C_F \frac{ 1 + (1-z)^2}{z}\red{.} 
\end{align}
It is clear from \red{Eq. (\ref{eq:4})} that when $Q_T$ is very small, the
logarithm $\ln^{(0,1)}(M^2_S/Q^2_T)/Q^2_T$ can become very large and 
perturbative expansion in $\alpha_s$ is no longer valid. The origin of
these large logarithms are due to long distance QCD effects: soft
and/or collinear radiation from initial-state partons. Thanks to QCD
factorization, the dynamics of soft and/or collinear radiation can be
well separated from the dynamics of UV physics. This is particular
useful for us, because we would like to perform a beam tagging study
in a way that does not rely too much on the underlying BSM models,
e.g., tree-level induced or loop-induced $S$ production. From
Eq. (\ref{eq:4}), one can also see that the leading logarithmic term differs
between $gg$-fusion cross section and $q\bar{q}$ annihilation cross
section, which is mainly due to the difference in the associated color
factor, $C_A=3$ versus $C_F = 4/3$. It is then expected that the
difference  can lead to different shape in the $Q_T$ spectrum. Since
the perturbative expansion of the $Q_T$ spectrum does not converge at
low $Q_T$, resummation of the large $Q_T$ logarithms is required
before one can assess the significance of the change in shape for the
$Q_T$ spectrum when switch between $gg$ fusion and $q\bar{q}$
annihilation. Fortunately, resumming the large logarithms due to small
transverse momentum have been studied since the early days of QCD~\cite{PHLTA.B79.269,NUPHA.B154.427,NUPHA.B193.381,NUPHA.B197.446,NUPHA.B250.199,hep-ph/0008184}. The
formalism developed in these pioneer works can be used in our 750 GeV
diphoton study with little change, thanks to the universality of
QCD at long distance. According to
the celebrated Collins-Soper-Sterman~(CSS) formula~\cite{NUPHA.B250.199}, the $Q_T$
distribution of the diphoton system can be written as an inverse Fourier transformation:
\begin{align}
  \label{eq:5}
  \frac{d\sigma^{(i)}}{dQ^2_T} = & \, \tau \hat{\sigma}^{(i)}_0 \int^\infty_0
  \frac{db}{2 }\, bJ_0(b Q_T) \sum_{a,b} \int^1_0 dx_1 \int^1_0 dx_2\, \delta( x_1 x_2  -\tau)
\nbrk
&\, \times \int^1_{x_1} \frac{d\xi_1}{\xi_1} C^{(i)}_{ia}\Big( \xi_1; \mu=\frac{b_0}{b}
  \Big)   f_{a/N_1}
  \Big(\frac{x_1}{\xi_1}; \mu = \frac{b_0}{b}\Big)  \int^1_{x_2} \frac{d\xi_2}{\xi_2}\, C^{(\bar{i})}_{\bar{i}b}\Big( \xi_2; \mu=\frac{b_0}{b} \Big) f_{b/N_2} \Big(\frac{x_2}{\xi_2}; \mu =
  \frac{b_0}{b}\Big)
\nbrk
&\,  \times  \exp\Bigg\{ -
  \int^{M^2_S}_{b^2_0/b^2} \frac{d \bar{\mu}^2}{\bar{\mu}^2} \Bigg[ \ln
  \frac{M^2_S}{\bar{\mu}^2} A^{(i)}[ \alpha_s(\bar{\mu})] + B^{(i)}[
  \alpha_s(\bar{\mu})] \Bigg] \Bigg\} 
    + \Big( i \leftrightarrow \bar{i} \Big)  +
  Y(Q^2_T,M^2_S,E_{\rm CM}^2)\red{,}
\end{align}
where $J_0(x)$ is the zeroth order Bessel function of the first kind,
$b_0 = 2e^{-\gamma_{\sss E}}$, $\gamma_{\sss E}=0.577216...$ is
Euler's constant. The summation of $a$ and $b$ are over different
parton species, $u,\bar{u}, d,\dots,g$. $A[ \alpha_s(\bar{\mu})]$ and $B[
\alpha_s(\bar{\mu})]$ are universal anomalous dimension whose
perturbative expansion can be written as
\begin{align}
  \label{eq:6}
  A^{(i)}[ \alpha_s(\mu)] = \sum_{n=1}^\infty
  \left(\frac{\alpha_s(\mu)}{4\pi} \right)^{n} A^{(i)}_n, \quad 
  B^{(i)}[ \alpha_s(\mu)] = \sum_{n=1}^\infty
  \left(\frac{\alpha_s(\mu)}{4\pi} \right)^{n} B^{(i)}_n\red{.}
\end{align}
In this work, we restrict ourselves to resummation of $Q_T$ logarithms
at Next-to-Leading Logarithmic (NLL) accuracy only, for which only
$A^{(i)}_1$, $A^{(i)}_2$ and $B^{(i)}_1$ are needed. They are given by~\cite{PHLTA.B112.66,PHLTA.B123.335,NUPHA.B244.337}
\begin{align}
  \label{eq:7}
  A^{(i)}_1 =&\, 4 C^{(i)}\red{,} 
\nbrk
  A^{(i)}_2 = &\,  4 C^{(i)}  \left(\left(\frac{67}{9}-\frac{\pi
   ^2}{3}\right) C_A-\frac{10
   N_f}{9}\right)\red{,}
\nbrk
  B^{(g)}_1 = &\, - \frac{22}{3} C_A + \frac{4}{3} N_f\red{,}
\nbrk
B^{(q)}_1 = &\, -6 C_F\red{,}
\end{align}
where $C^{(g)} = C_A$, $C^{(q)}=  C_F$.
The function $C^{(i)}_{ij}(x;\mu)$ is the hard collinear factor. For
NLL resummation, we only need their LO expression:
\begin{align}
  \label{eq:8}
  C^{(i)}_{ij} (x;\mu) = \delta_{ij} \delta(1-x)\red{.}
\end{align}
$Y(Q^2_T,\tau)$ denotes the terms which are not enhanced by
$\ln(M^2_S/Q^2_T)$. They can be computed using naive expansion in
$\alpha_s$. Sometimes they could have large impact at large $Q_T$. But
in the region we are interested in, they can be safely neglected. 
Note that in Eq. (\ref{eq:5}), when $b$ is very large, the integral for
$\bar{\mu}$ in the exponent would hit Landau pole, where
$\alpha_s(\bar{\mu})$ diverges. The existence of Landau pole at small
$\bar{\mu}$ indicates the onset of non-perturbative physics in that
region, and appropriate prescription to deal with the Landau pole is
needed, see, e.g., Refs.~\cite{NUPHA.B250.199,hep-ph/9311341,hep-ph/0012348,hep-ph/0508068}.  
We emphasize that the CSS formula is quite general and doesn't depend
too much on the UV dynamics of the underlying process. Remarkably, at NLL level, all the process
dependent information have been encoded in the tree partonic cross
section $\hat{\sigma}^{(i)}_0$, and in the label $(i)$ for various
dimension and collinear factor. Thus, we expect that the statement we
make from the $Q_T$ spectrum is rather model independent. 

To quantify the discussion above, we calculate the $Q_T$ spectrum of
the 750 GeV diphoton system numerically for $13$ TeV LHC. Thanks to
the previous QCD studies, several
public computer codes are available which implement the resummation of
transverse momentum logarithms for Drell-Yan and Higgs production,
both in the QCD framework and in the Soft-Collinear Effective theory
framework~\cite{hep-ph/0005275,hep-ph/0011336,hep-ph/0109045,hep-ph/0202088}. Resummation of $Q_T$ for 750 GeV diphoton resonance can be
easily accomplished by modifying those existing codes. Specifically,
we modify \texttt{HqT}, which is based on the work of Refs.~\cite{hep-ph/0302104,hep-ph/0508068,1109.2109,1007.2351}, and
\texttt{CuTe}, which is based on the work of refs.~\cite{1007.4005,1212.2621}, to calculate the
transverse momentum spectrum of the hypothetical 750 GeV resonance. In
\texttt{HqT}, Landau pole is avoided by deforming the $b$-space
integral off the real axis slightly. While in \texttt{CuTe}, the
Landau pole is avoided by imposing a cutoff for the $\bar{\mu}$ integral at
very small value. In both calculations, we use the five-flavor
scheme, namely the bottom quark is treated as a massless
parton in the PDFs.

\begin{figure}[h]
  \begin{center}
  \includegraphics[width=0.45\textwidth]{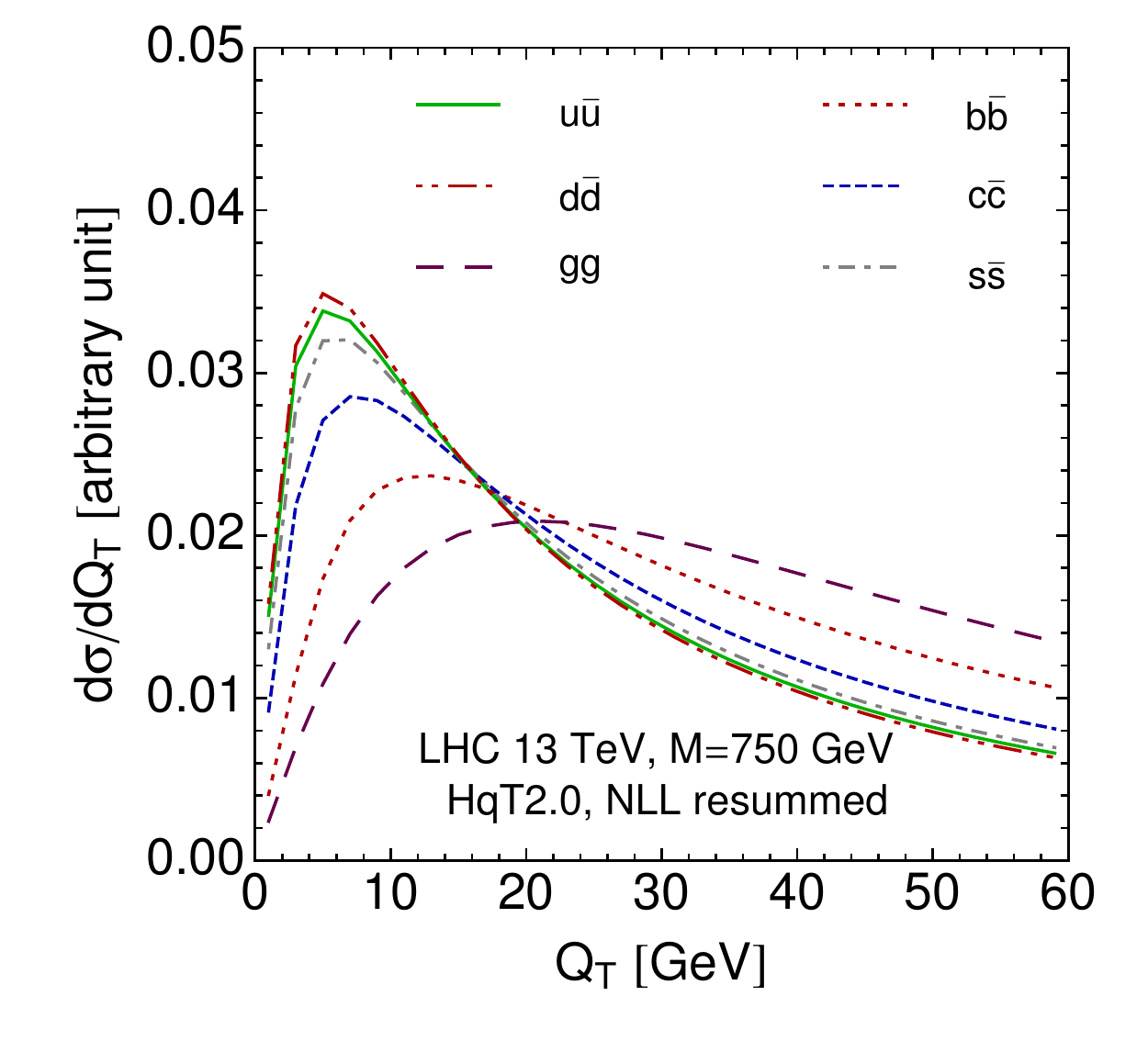}
\hspace{1cm}
  \includegraphics[width=0.45\textwidth]{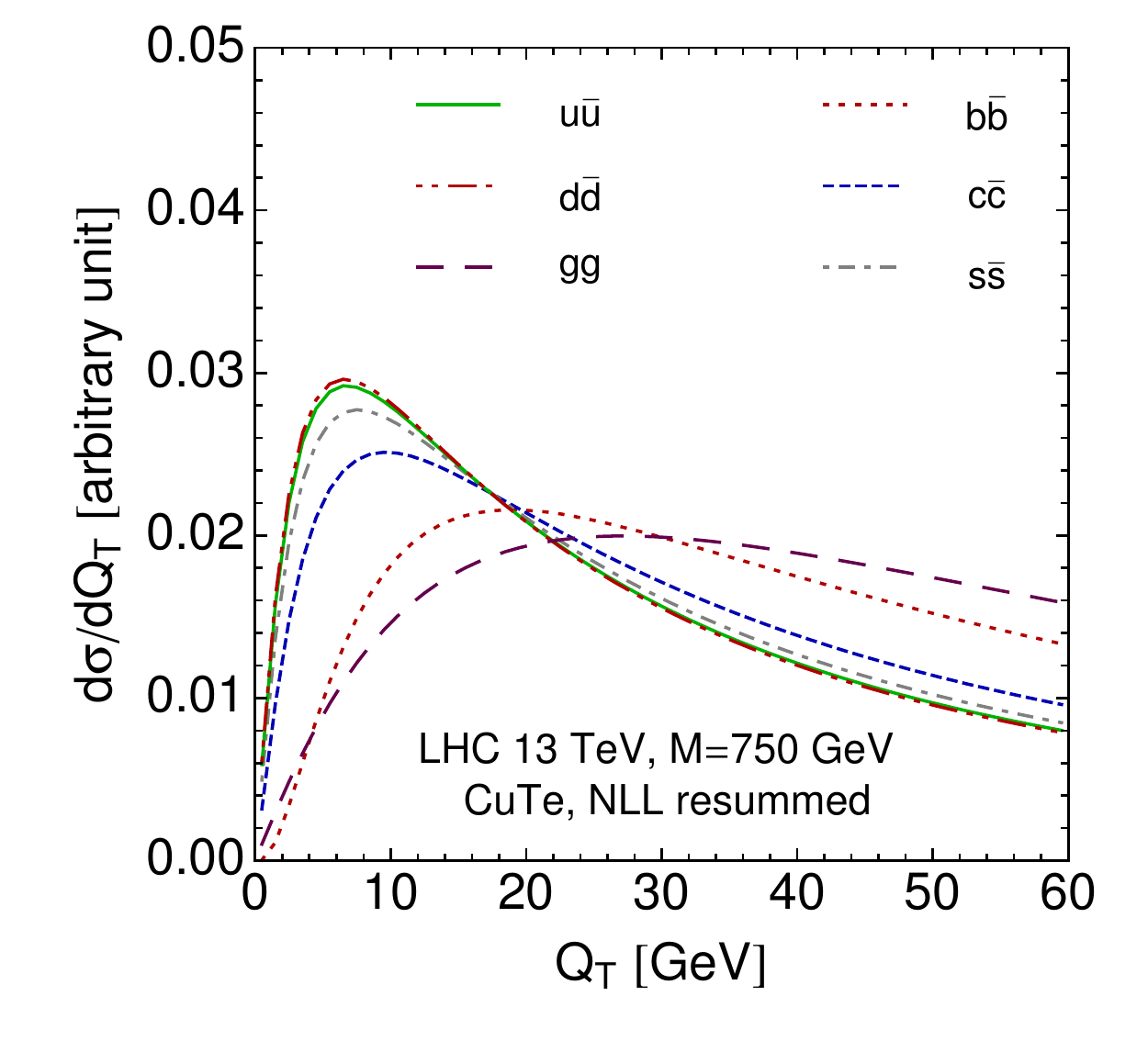}
  \end{center}
  \vspace{-2ex}
  \caption{\label{fig:qt}
  $Q_T$ distribution at small transverse momentum at NLL accuracy for
  the resonace production initiated by different parton flavors. The two
  plots show results obtained from two public codes, \texttt{HqT} and \texttt{CuTe}.}
\end{figure}

We
calculate the $Q_T$ spectrum by turning on the coupling of the
diphoton resonance
with each individual parton flavor at one time. The differential
distribution is plotted in FIG.~\ref{fig:qt} for
results from the two codes mentioned above at NLL resummed
accuracy. Comparing the distributions for production initiated
by different parton combinations, the shapes are mostly driven by
two factors: a) the color factor in Sudakov exponent, $C_A$ for gluon
versus $C_F$ for quarks; b) the evolution of PDFs. For light-quark
contributions, which includes up, down, strange, and charm quark, the peak position stay at low values, less than 10 GeV
in general. For bottom-quark case,
the distributions are broader and shift to higher $Q_T$. The reason
for the rightward shift of the bottom contribution comparing to the
light-quark contribution is as follows. For the formal treatment of the
quark contribution in the CSS formula, Eq. (\ref{eq:5}), there are no
essential difference between light quark and bottom quark. The only
difference comes from their PDFs, which are evaluated at the scale
$b_0/b$, the Fourier conjugate of $Q_T$. While the DGLAP evolution for light quark and bottom quark
are the same in the five flavor scheme, the boundary conditions for
these PDFs differ. For bottom quark, the threshold of the
corresponding PDF lies around $m_b \sim 4.2$ GeV, below which the PDF
vanishes. On the other hand, the threshold of the light quark PDFs
lies around much lower values than the bottom quark one. It thus
indicates that the Sudakov peak for bottom-quark contribution has to show up at larger value of
$Q_T$ in order to accommodate the fact that its threshold is higher.
For the gluon contribution, the shape of the $Q_T$ spectrum is further
broadened, and has the largest value for the peak position. This is
mainly due to the difference in color factor. In the gluon case, the
Sudakov exponent has a stronger suppression effects because $C_A \sim 2.25\,
C_F$. We have checked that if we naively change the color factor from
$C_A$ to $C_F$ for the gluon contribution, its peak position move to a much
lower value.
From FIG.~\ref{fig:qt}, we can see that the results
from the two codes used for the calculation are similar, although
they have different framework for resummation, and different treatment
of Landau pole. The major difference comes from the bottom quark
contribution, where the peak position differ by about $5$ GeV. This is
mainly due to different ways in the two codes to avoid Landau pole.
Because of the large mass of the resonance, non-perturbative effects
are less pronounced as comparing with the $W$, $Z$ boson production in
the SM, as we checked by varying the non-perturbative parameter available in
\texttt{HqT} and \texttt{CuTe}. Also, for the same reason, the
subleading terms in $Q_T$ are small in the region we plot.

\begin{table}[!h]
\centering
\begin{tabular}{ccccccc} 
\hline \hline 
$R$, NLL & $b\bar b$ & $c\bar c$ & $s\bar s$ & $u\bar u$ &  $d\bar d$ & $gg$ \\
\hline \hline 
\texttt{HqT} & 0.95 & 0.68 & 0.58 & 0.55 & 0.53 & 1.32  \\
\texttt{CuTe} & 1.32 & 0.82 & 0.70 & 0.66 & 0.65 & 1.52 \\
\hline \hline 
\end{tabular}
\caption{Ratio $R$ for a 750 GeV resonance produced at 13 TeV LHC, initiated by
different parton flavors as predicted by two resummation codes, \texttt{HqT} and \texttt{CuTe}.}
\label{tab:r}
\end{table}
Ideally, a detailed comparison of the normalized $Q_T$ distribution
predicted by QCD factorization and the LHC data for the
hypothetical resonance would provide most information about the beam
tagging problem from $Q_T$ spectrum. In reality, this is very
difficult due to the limited statistics and experimental
uncertainties in measuring the photon transverse momentum. 
To simplify the analysis, we introduce a ratio $R$, which is defined
as the cross section in $Q_T$ bin of $[\Delta_T,2 \Delta_T]$ to the one
in $Q_T$ bin of $[0,\Delta_T]$. The optimal choice for $\Delta_T$
differs for different center of mass energy and different resonance
mass. In our current case, we choose $\Delta_T=20$ GeV.  The results
for the ratio are listed in TABLE.~\ref{tab:r}
based on curves shown in FIG.~\ref{fig:qt} for the two codes and
various parton flavors. We can see a clear distinction for production
initiated by light quarks, which favor a value of $R$ lower than $1$,
and production initiated by gluon, which favors a value of $R$ larger
than $1$. As noted above, prediction for bottom-quark initiated
production are quite different, indicating a larger theoretical uncertainty
in the resummation treatment of heavy-quark induced diphoton
production. This uncertainty prevents us from distinguishing 
it from gluon initiated case. The uncertainty might be reduced
if the calculation is extended to NNLL level consistently, or using
four-flavor scheme for the PDFs, which are beyond the scope of this
work. We have also checked the theoretical uncertainties from other sources, e.g., PDFs and
power corrections which are at a few percent level and can be neglected safely.   

\subsection{Diphoton with additional $b$-jet}
\label{sec:diphoton-bjet}
In the previous two sections, we have shown that by measuring the rapidity
and transverse-momentum distribution of the diphoton system, it is
possible to distinguish the valence-quark induced diphoton production
from sea-quark/gluon induced diphoton production, and light-quark induced
diphoton production from gluon induced diphoton production.
In this section, we focus on the remaining two production scenarios. In the 
first scenario (gg), the new scalar resonance is produced via 
the gluon fusion process. In the second scenario (bb), the scalar
resonance is produced via $b\bar b$ initial state. 
We will show that a 99.7\% C.L. distinguish
can be reached with less than 10fb$^{-1}$ integrated luminosity
at 13 TeV LHC. This means if the 750GeV excess is 
indeed a new resonance, we do not need to wait for long 
to know its production mechanism.

\begin{figure}[h]
  \begin{center}
  \includegraphics[width=0.55\textwidth]{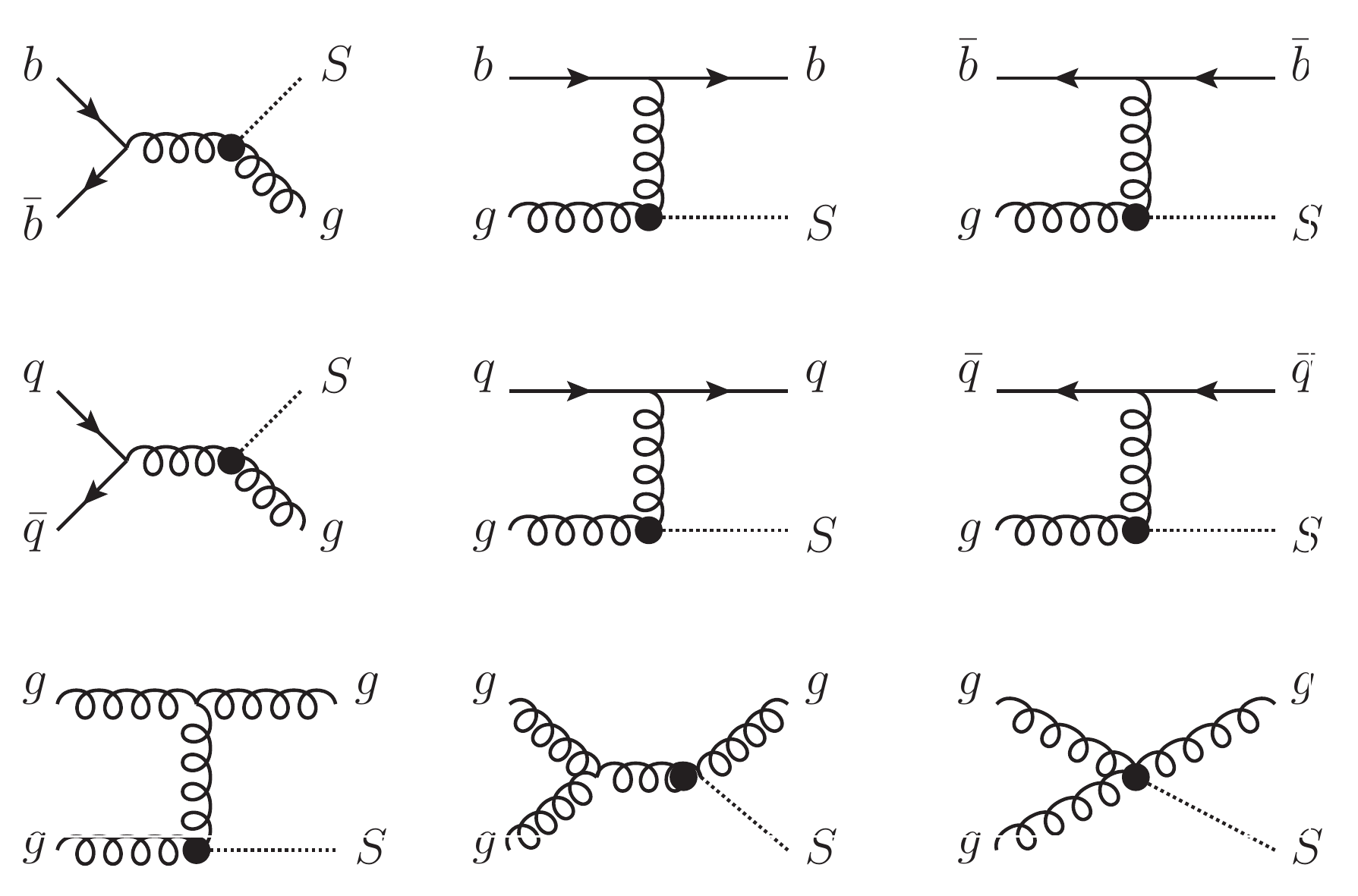}
  \end{center}
  \vspace{-2ex}
  \caption{\label{fig:gg_1b}
   The Feynman diagrams of the resonance with one jet 
   production process in gg scenario.}
\end{figure}

In the gg scenario, the dominant production mode of the new
resonance is gluon fusion process. With the initial state radiation (ISR)
effect, there are additional jets in the final state. The 
Feynman diagrams for jet production at LO in QCD are shown in
FIG. \ref{fig:gg_1b}. Since in the small-$x$ region the gluon 
PDF is much larger than other partons, it is easy to know that
most of the ISR jets are gluon and light (especially $u$ and $d$)
quarks. The $b$-jet fraction in the ISR jets is highly
suppressed by the small bottom-quark PDF. Thus we expect that
there is very seldom hard $b$-jet in the ISR
jets.

\begin{figure}[h]
  \begin{center}
  \includegraphics[width=0.7\textwidth]{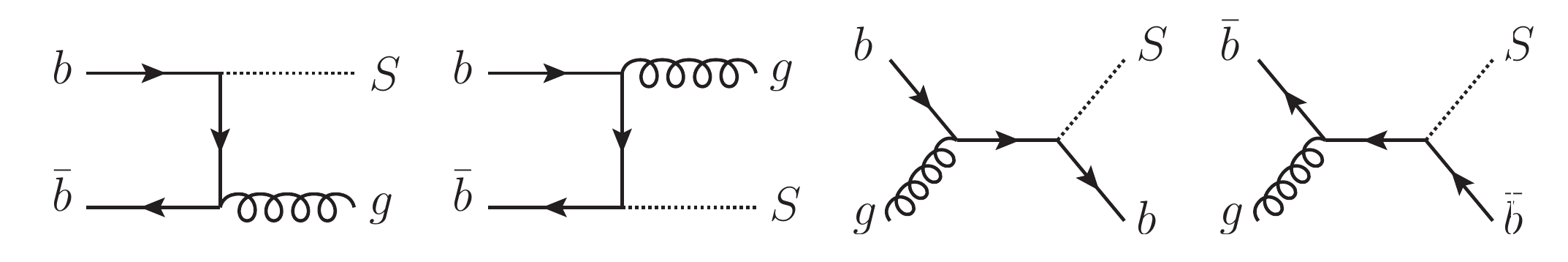}
  \end{center}
  \vspace{-2ex}
  \caption{\label{fig:bb_1b}
   The Feynman diagrams of the resonance with one jet production
   process. In this scenario, the new resonance is produced via the 
   $b\bar b$ initial state at the LHC.}
\end{figure}

In the bb scenario, we show the 
Feynman diagrams for jet production at LO in QCD in
FIG. \ref{fig:bb_1b}. The large gluon PDF induces
a lot of $b$-jets from the $gb(\bar b)$ initial state processes. 
The $b$-jet fraction in the ISR jets then should be significant and 
can be tagged at the LHC Run 2.

\begin{figure}[htb!]
  \includegraphics[width=0.45\textwidth]{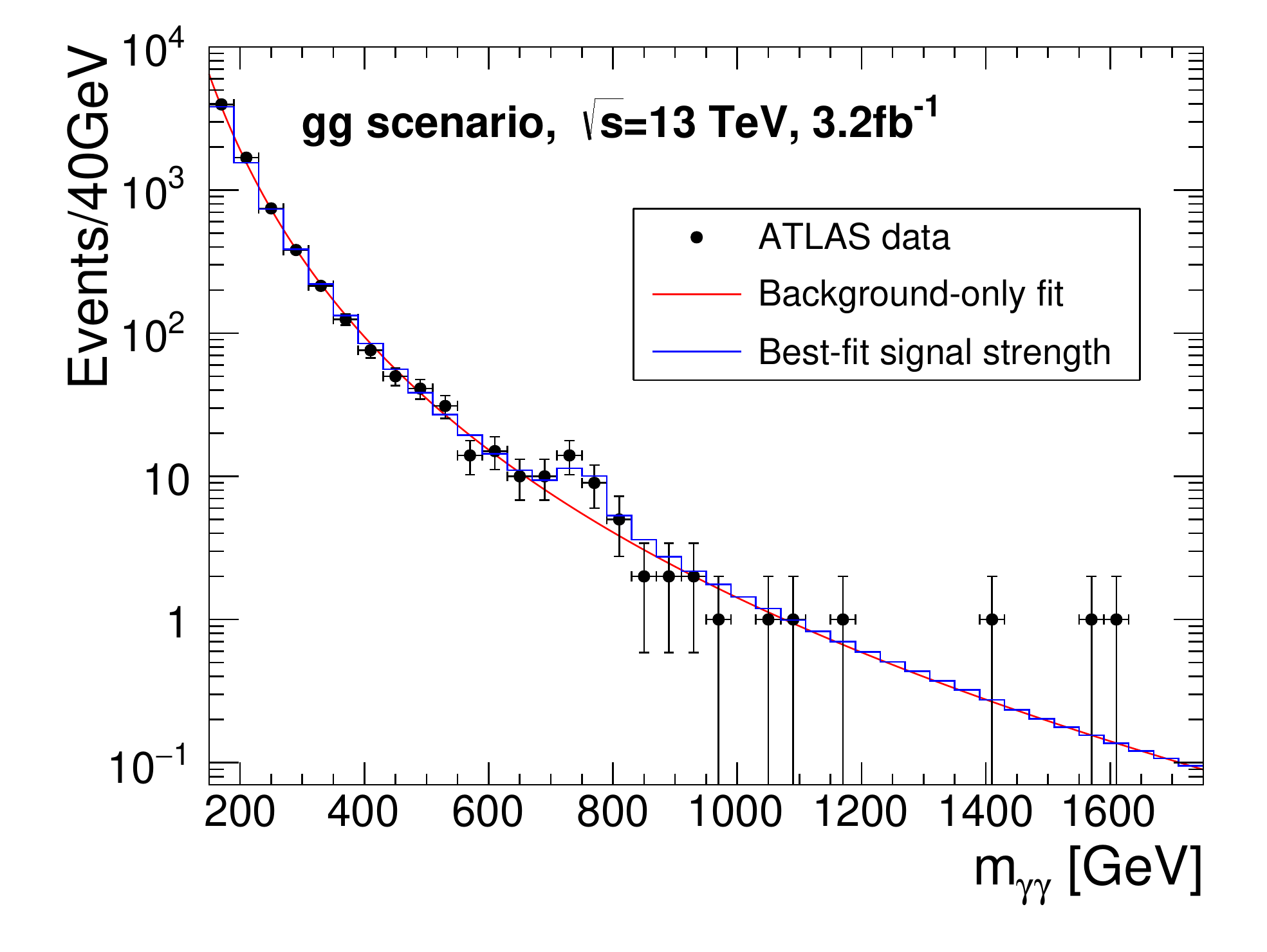}
  \includegraphics[width=0.45\textwidth]{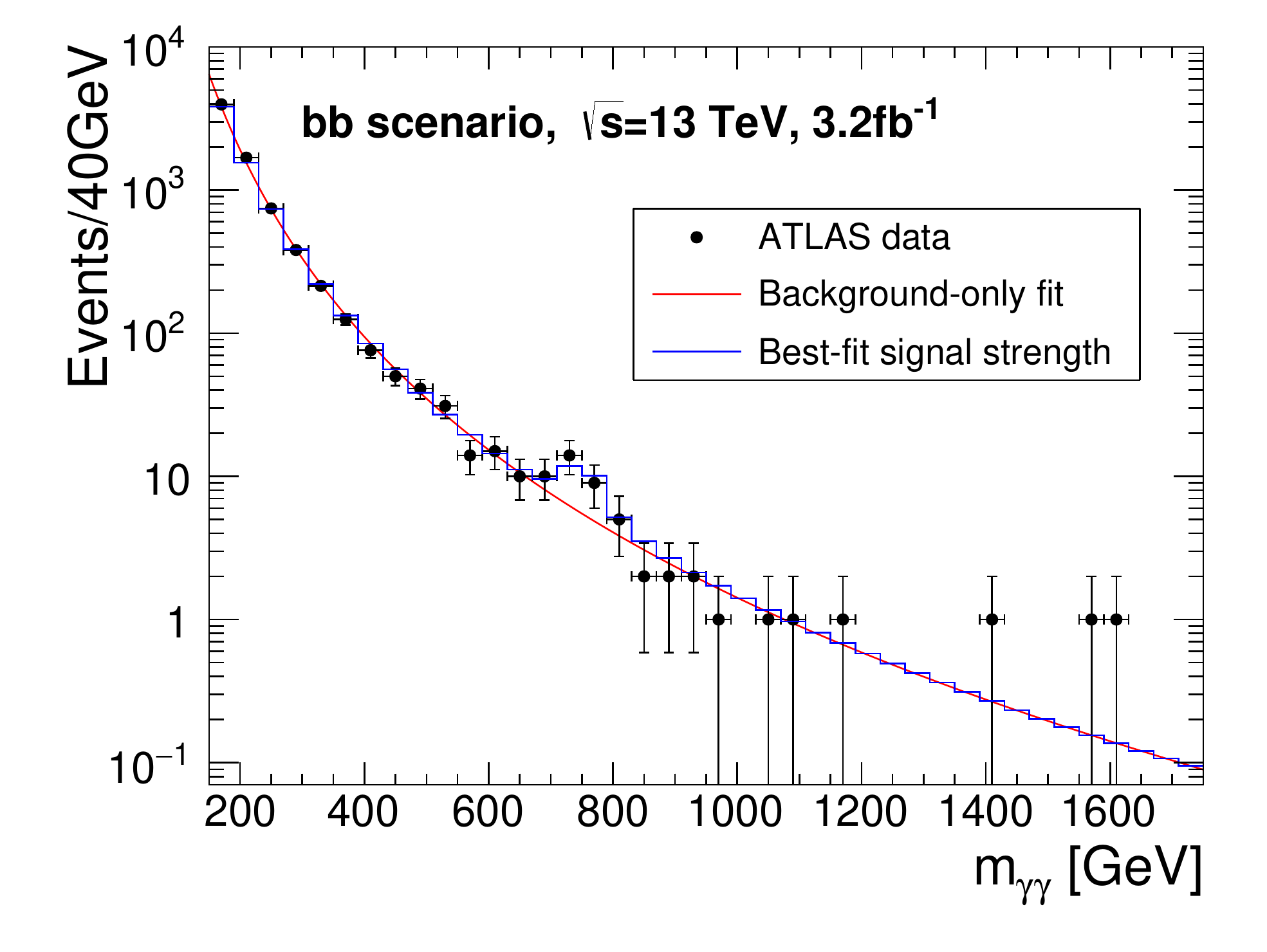}
  \caption{\label{fig:maa}
   The best-fit results of the new physics models. Upper panel: the gg 
   scenario; Lower panel: the bb scenario.}
\end{figure}

For a simple estimation, we generate parton level 
signal events with MadGraph5 \cite{Alwall:2014hca} 
and CT14llo PDF (5 flavor scheme) \cite{Dulat:2015mca}. 
The signal events are showered using Pythia6.4 \cite{Sjostrand:2006za} 
with Tune Z2 parameter \cite{Field:2011iq}. The detector effect
is simulated using DELPHES 3 \cite{deFavereau:2013fsa,Cacciari:2011ma}.
The $b$-tagging efficiency is tuned to be consistent 
with the distribution shown in Ref \cite{ATL-PHYS-PUB-2015-022}.  
For the signal strength, we scale the inclusive signal events
(with MLM matching scheme) to fit the current data \cite{ATLAS-CONF-2015-081,
CMS-PAS-EXO-15-004} (in this work, we only fit the data
from the ATLAS collaboration).
We require the photon to satisfy
\begin{equation}
|\eta|<1.37,~{\text{or}}~1.52<|\eta|<2.37.
\end{equation}
The transverse energy of the leading (subleading) 
photon should be larger than 40 (30) GeV. The leading 
and subleading photon candidates are then
required to satisfy the conditions 
\begin{equation}
\frac{E_{\text{T}}^{\gamma_1}}{m_{\gamma\gamma}}>0.4,~
\frac{E_{\text{T}}^{\gamma_2}}{m_{\gamma\gamma}}>0.3.
\end{equation}
The inclusive diphoton spectrum is estimated with 
\begin{equation}
0.0255\left[1-\left(\frac{m_{\gamma\gamma}}{13000{\text{GeV}}}\right)
^{1/45}\right]^{3.38248}\left(\frac{m_{\gamma\gamma}}{13000{\text{GeV}}}\right)^{-3.49062}
{\text{fb/GeV}}.
\label{eq:bkgd}
\end{equation}
We solve the best-fit signal strength $\mu$ by maximizing \cite{Read:2002hq,Cowan:2010js}
\begin{equation}
\sqrt{-2\ln\left[\frac{L\left(\{b\}|\{n\}\right)}{L\left(\mu \{s\}+\{b\}|\{n\}\right)}\right]},
\end{equation}
where the likelihood function is defined by
\begin{equation}
L\left(\{x\}|\{n\}\right)\equiv\prod_{i}\frac{x_i^{n_i}\exp\left(-x_i\right)}{\Gamma\left(n_i+1\right)}.
\end{equation}
Both the gg and the bb scenario give 3 $\sigma$ discovery significance.
The best-fit results are shown in FIG. \ref{fig:maa}.

\begin{figure}[htb!]
  \begin{center}
  \includegraphics[width=0.55\textwidth]{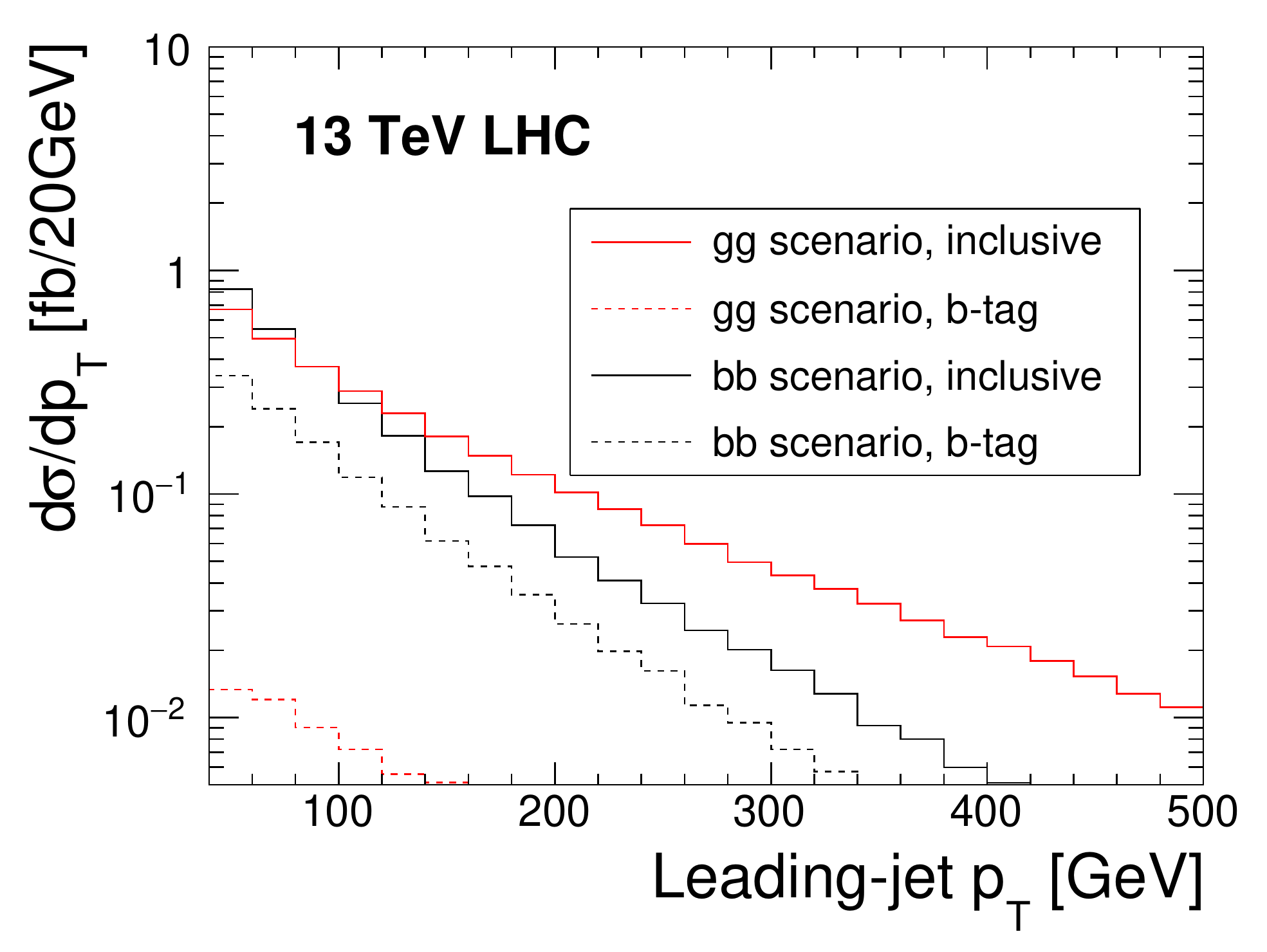}
  \end{center}
  \vspace{-2ex}
  \caption{\label{fig:bjetpt}
   The leading-($b$-)jet transverse momentum distribution.
   The distributions are normalized by the inclusive event number
   with additional jets.}
\end{figure}

After rescaling the inclusive cross section to the best-fit value,
we investigate the events with at least one hard jet in the final
state.
Additional jets in the final state are reconstructed with anti-$k_{\text{T}}$
jet algorithm with $R=0.4$. The hard jets must satisfy
\begin{equation}
|\eta|<2.5,~p_{\text{T}}>40{\text{GeV}}.
\end{equation}
To suppress the SM background, we add diphoton invariant mass
cut $|m_{\gamma\gamma}-750{\text{GeV}}|<150{\text{GeV}}$.
The leading jet transverse momentum distributions with and without 
$b$-tagging are shown in FIG. \ref{fig:bjetpt}. 
At 13 TeV LHC, there will be 0.0768fb $b$-jet event
in 3.115fb signal events with at least one additional jet in gg scenario,
and 1.212fb $b$-jet event
in 2.717fb signal events with at least one additional jet in bb scenario.

To give an estimation of the possibility of distinguishing the 
two production scenarios, we also need to simulate the SM
backgrounds. There are lots of theoretical uncertainties. And 
only a data driven estimation of the backgrounds is reliable. 
In this work, we make a simple estimation by rescaling the 
current background with luminosity. Thus we only need to 
calculate the fraction of the background events with 
additional hard $b$-jet. The most important SM backgrounds 
are the irreducible $\gamma\gamma$ process and the 
reducible $\gamma j$ and $jj$ processes with one or more
jets faked to be photon in the detector. With the mass window 
cut, we count the fraction of events with at least one additional 
hard jet ($N_{+j}/N_{incl}$), and the fraction of these events whose leading jet is 
tagged as a $b$-jet ($N_{+b}/N_{+j}$). Since the cut on the first 
and the second photon transverse energy are asymmetric, there
are a lot of events which pass the cuts with additional jets from 
the ISR. The results are shown in Table. \ref{tab:bkgd}.
With the data driven background formula Eq. (\ref{eq:bkgd}), the 
total background cross section in $600{\text{GeV}}<m_{\gamma\gamma}<900
{\text{GeV}}$ is $15.32$fb.
\begin{table*}[!htb]
\caption{The fraction of the background events with at least one
additional hard jet. And the fraction of the events with the leading 
jet is tagged as a $b$-jet in these events. In the last line, we show
the $N_{+b}$ event number with the assumption that all background
events are from the corresponding process.}
\label{tab:bkgd}
\begin{center}
\begin{tabular}{cccc}
\hline\hline
Background & $\gamma\gamma$ & $\gamma j$ & $jj$\\
\hline\hline
$N_{+j}/N_{incl}$ & $47.1\%$ & $66.3\%$ & $64.5\%$ \\
$N_{+b}/N_{+j}$ & $1.85\%$ & $2.63\%$ & $5.03\%$ \\
~~~$N_{+b}/N_{incl}$~~~ & ~~~$0.871\%$~~~ & ~~~
$1.74\%$~~~ & ~~~$3.24\%$~~~ \\
$N_{+b}$ (fb) & $0.133$ & $0.267$ & $0.497$ \\
\hline\hline\\
\end{tabular}
\end{center}
\end{table*}

\begin{figure}[htb!]
  \begin{center}
  \includegraphics[width=0.55\textwidth]{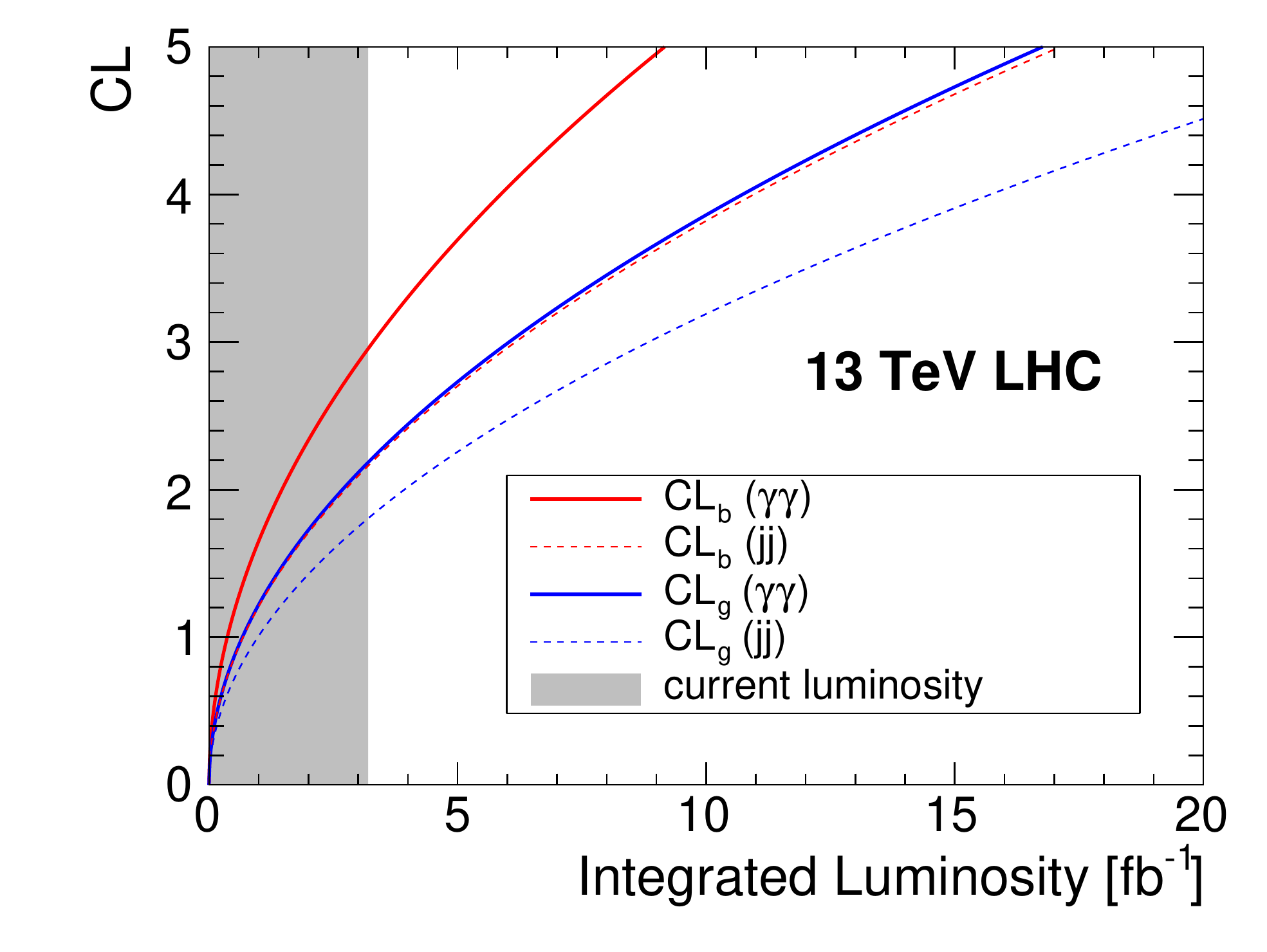}
  \end{center}
  \vspace{-2ex}
  \caption{\label{fig:bdis}
   The ability of distinguishing the gg (bb) scenario from the bb (gg)
scenario. The solid lines are with the assumption that all of the 
background events are from the irreducible $\gamma\gamma$ background.
The dashed lines are with the assumption that the all of the 
background events are from the reducible $jj$ background with two 
jets are faked as photon.}
\end{figure}

Since the background cross section is small, we estimate 
the ability of distinguishing the gg (bb) scenario from the bb (gg)
scenario with \cite{Read:2002hq,Cowan:2010js}
\begin{equation}
{\text{CL}}_{g}\equiv \sqrt{-2\log\left[\frac{L\left(s_b+n_b|s_g+n_b\right)}
{L\left(s_g+n_b|s_g+n_b\right)}\right]}~~
\left({\text{CL}}_{b}\equiv \sqrt{-2\log\left[\frac{L\left(s_g+n_b|s_b+n_b\right)}
{L\left(s_b+n_b|s_b+n_b\right)}\right]}\right),
\end{equation}
where $s_b,s_g$ and $n_b$ are the event numbers with the leading additional
jet tagged as a $b$-jet in the scenario bb, scenario gg and the SM background. 
In FIG. \ref{fig:bdis}, we show the distinguishing abilities versus the integrated luminosity 
of the 13 TeV LHC. It is shown clearly in this figure that, even with the most conservative 
assumption (all background events are from the $jj$ process), one can distinguish 
the gg scenario from the bb scenario with 8.8fb$^{-1}$ integrated luminosity,
and distinguish the bb scenario from the gg scenario with 6.2fb$^{-1}$ integrated luminosity
at 13 TeV LHC. If the SM background are (a MC simulation will support this assumption)
$\gamma\gamma$ process dominant, one can distinguish 
the gg scenario from the bb scenario with 6.0fb$^{-1}$ integrated luminosity,
and distinguish the bb scenario from the gg scenario with 3.3fb$^{-1}$ integrated luminosity
at 13 TeV LHC. 

\section{Summary and conclusion}
\label{sec:summary-conclusion}

Recently, an intriguing excess in the diphoton events has been
reported both by the ATLAS and CMS collaboration. The local
significance is \red{3.9$\sigma$ from ATLAS and 2.6$\sigma$ from
CMS.} After taking into account the look-elsewhere effect, the
significance reduces to \red{2.3$\sigma$ from ATLAS and 1.2$\sigma$ from
CMS.} Although the current experimental status
is far from conclusive, a large number of BSM scenarios have been
explored to explain the diphoton excess. A significant number of these
BSM models contain a scalar resonance produced from hadron-hadron
collision and subsequently decay to diphoton system, whose mass is
around 750 GeV. In this work, we investigated whether the hadronic production
mechanism for the  hypothetical new scalar resonance can be
identified. That is, is it mainly produced from
$gg$ fusion or $q\bar{q}$ annihilation. We dubbed this question the
quark and gluon beam tagging problem. We expect that a successful
solution to this problem will play a key role in unraveling the mystery of
the 750 GeV diphoton excess. We have performed a model independent studied of this problem  by considering a set of effective operator\red{s}
between the hypothetical resonance and gluon or quark. We discuss
several differential distributions relevant for the determination of
initial constituent for the 750 GeV excess. We concentrate on those
distributions which are more sensitive to QCD dynamics at long
distance, and thus less model dependent. To that end, we explored
three different but complementary observables for beam
tagging. Firstly, we calculated the rapidity distribution of the
diphoton system, and found that it is helpful for distinguishing
valence quark induced production from gluon or sea quark induced
production. The main reason is that the PDFs for $u$ and $d$ quark are
much larger at large $x$, comparing to $\bar{u}$ and $\bar{d}$
quark. Secondly, we calculated the transverse-momentum spectrum of the
diphoton system and focus on the small $Q_T$ region, where a Sudakov
peak is formed due to multiple soft and/or collinear radiation from
initial state. We found that a clear distinction for the light quark
induced production from the gluon or $b$-quark induced production can
be achieved. This is mainly due to the difference in the effective
strength of initial state bremsstrahlung: for light quark it is $C_F
\alpha_s = \frac{4}{3} \alpha_s$, while for gluon it is $C_A \alpha_s
= 3 \alpha_s$. Such difference leads to a notable shift of the peak
towards larger $Q_T$, as well as a much broader peak. For $b$ quark induced
production, the difference in the peak structure from gluon induced
production is less pronounced, due to the large $b$ quark mass and
uncertainty associated with $Q_T$ resummation in five flavor scheme
versus four flavor scheme. Thirdly, in order to distinguish the gluon
induced production from $b$ quark induced production, we calculated
the diphoton plus jet production with a $b$ tagging on the leading
jet in five flavor scheme. We find that an additional $b$ jet is more
favored in $b$ quark induced production than in gluon induced
production. Combining the knowledge gain from all there observables,
we find that the perspective for identifying the exact production
mechanism for the hypothetical diphoton resonance is promising,
though detailed work is needed in order to further understand the
theory and experimental uncertainties of our methods, which we leave
for future work. Lastly, we emphasize that although the current work
is mainly motivated by the diphoton excess recently reported by the ATLAS
and CMS collaboration, the problem we proposed and the methods we
suggested are useful and interesting in itself even the excess
disappears after more data is collected.

\section{acknowledgments}
We thank Yotam Soreq and Wei Xue for helpful conversations.
The work of H.Z. is supported by the U.S. DOE under Contract No. 
DE-SC0011702. The work of H.X.Z. is supported by the U.S. Department
of Energy under grant Contract Number DE-SC0011090. 
Work at ANL is supported in part by the U.S. Department of Energy
under Contract No. DE-AC02-06CH11357. \red{H.Z. is pleased to recognize 
the hospitality of the service offered by the Amtrak California Zephyr train.}

\appendix
\bibliographystyle{apsrev}
\bibliography{GZZ_diphoton}

\begin{thebibliography}{138}
\expandafter\ifx\csname natexlab\endcsname\relax\def\natexlab#1{#1}\fi
\expandafter\ifx\csname bibnamefont\endcsname\relax
  \def\bibnamefont#1{#1}\fi
\expandafter\ifx\csname bibfnamefont\endcsname\relax
  \def\bibfnamefont#1{#1}\fi
\expandafter\ifx\csname citenamefont\endcsname\relax
  \def\citenamefont#1{#1}\fi
\expandafter\ifx\csname url\endcsname\relax
  \def\url#1{\texttt{#1}}\fi
\expandafter\ifx\csname urlprefix\endcsname\relax\def\urlprefix{URL }\fi
\providecommand{\bibinfo}[2]{#2}
\providecommand{\eprint}[2][]{\url{#2}}

\bibitem[{ATL(2015{\natexlab{a}})}]{ATLAS-CONF-2015-081}
\bibinfo{type}{Tech. Rep.} \bibinfo{number}{ATLAS-CONF-2015-081},
  \bibinfo{institution}{CERN}, \bibinfo{address}{Geneva}
  (\bibinfo{year}{2015}{\natexlab{a}}),
  \urlprefix\url{http://cds.cern.ch/record/2114853}.

\bibitem[{CMS(2015)}]{CMS-PAS-EXO-15-004}
\bibinfo{type}{Tech. Rep.} \bibinfo{number}{CMS-PAS-EXO-15-004},
  \bibinfo{institution}{CERN}, \bibinfo{address}{Geneva}
  (\bibinfo{year}{2015}), \urlprefix\url{http://cds.cern.ch/record/2114808}.

\bibitem[{\citenamefont{Harigaya and Nomura}(2015)}]{Harigaya:2015ezk}
\bibinfo{author}{\bibfnamefont{K.}~\bibnamefont{Harigaya}} \bibnamefont{and}
  \bibinfo{author}{\bibfnamefont{Y.}~\bibnamefont{Nomura}}
  (\bibinfo{year}{2015}), \eprint{arXiv:1512.04850}.

\bibitem[{\citenamefont{Mambrini et~al.}(2015)\citenamefont{Mambrini, Arcadi,
  and Djouadi}}]{Mambrini:2015wyu}
\bibinfo{author}{\bibfnamefont{Y.}~\bibnamefont{Mambrini}},
  \bibinfo{author}{\bibfnamefont{G.}~\bibnamefont{Arcadi}}, \bibnamefont{and}
  \bibinfo{author}{\bibfnamefont{A.}~\bibnamefont{Djouadi}}
  (\bibinfo{year}{2015}), \eprint{arXiv:1512.04913}.

\bibitem[{\citenamefont{Backovic et~al.}(2015)\citenamefont{Backovic, Mariotti,
  and Redigolo}}]{Backovic:2015fnp}
\bibinfo{author}{\bibfnamefont{M.}~\bibnamefont{Backovic}},
  \bibinfo{author}{\bibfnamefont{A.}~\bibnamefont{Mariotti}}, \bibnamefont{and}
  \bibinfo{author}{\bibfnamefont{D.}~\bibnamefont{Redigolo}}
  (\bibinfo{year}{2015}), \eprint{arXiv:1512.04917}.

\bibitem[{\citenamefont{Angelescu et~al.}(2015)\citenamefont{Angelescu,
  Djouadi, and Moreau}}]{Angelescu:2015uiz}
\bibinfo{author}{\bibfnamefont{A.}~\bibnamefont{Angelescu}},
  \bibinfo{author}{\bibfnamefont{A.}~\bibnamefont{Djouadi}}, \bibnamefont{and}
  \bibinfo{author}{\bibfnamefont{G.}~\bibnamefont{Moreau}}
  (\bibinfo{year}{2015}), \eprint{arXiv:1512.04921}.

\bibitem[{\citenamefont{Nakai et~al.}(2015)\citenamefont{Nakai, Sato, and
  Tobioka}}]{Nakai:2015ptz}
\bibinfo{author}{\bibfnamefont{Y.}~\bibnamefont{Nakai}},
  \bibinfo{author}{\bibfnamefont{R.}~\bibnamefont{Sato}}, \bibnamefont{and}
  \bibinfo{author}{\bibfnamefont{K.}~\bibnamefont{Tobioka}}
  (\bibinfo{year}{2015}), \eprint{arXiv:1512.04924}.

\bibitem[{\citenamefont{Knapen et~al.}(2015)\citenamefont{Knapen, Melia,
  Papucci, and Zurek}}]{Knapen:2015dap}
\bibinfo{author}{\bibfnamefont{S.}~\bibnamefont{Knapen}},
  \bibinfo{author}{\bibfnamefont{T.}~\bibnamefont{Melia}},
  \bibinfo{author}{\bibfnamefont{M.}~\bibnamefont{Papucci}}, \bibnamefont{and}
  \bibinfo{author}{\bibfnamefont{K.}~\bibnamefont{Zurek}}
  (\bibinfo{year}{2015}), \eprint{arXiv:1512.04928}.

\bibitem[{\citenamefont{Buttazzo et~al.}(2015)\citenamefont{Buttazzo, Greljo,
  and Marzocca}}]{Buttazzo:2015txu}
\bibinfo{author}{\bibfnamefont{D.}~\bibnamefont{Buttazzo}},
  \bibinfo{author}{\bibfnamefont{A.}~\bibnamefont{Greljo}}, \bibnamefont{and}
  \bibinfo{author}{\bibfnamefont{D.}~\bibnamefont{Marzocca}}
  (\bibinfo{year}{2015}), \eprint{arXiv:1512.04929}.

\bibitem[{\citenamefont{Pilaftsis}(2015)}]{Pilaftsis:2015ycr}
\bibinfo{author}{\bibfnamefont{A.}~\bibnamefont{Pilaftsis}}
  (\bibinfo{year}{2015}), \eprint{arXiv:1512.04931}.

\bibitem[{\citenamefont{Franceschini et~al.}(2015)\citenamefont{Franceschini,
  Giudice, Kamenik, McCullough, Pomarol, Rattazzi, Redi, Riva, Strumia, and
  Torre}}]{Franceschini:2015kwy}
\bibinfo{author}{\bibfnamefont{R.}~\bibnamefont{Franceschini}},
  \bibinfo{author}{\bibfnamefont{G.~F.} \bibnamefont{Giudice}},
  \bibinfo{author}{\bibfnamefont{J.~F.} \bibnamefont{Kamenik}},
  \bibinfo{author}{\bibfnamefont{M.}~\bibnamefont{McCullough}},
  \bibinfo{author}{\bibfnamefont{A.}~\bibnamefont{Pomarol}},
  \bibinfo{author}{\bibfnamefont{R.}~\bibnamefont{Rattazzi}},
  \bibinfo{author}{\bibfnamefont{M.}~\bibnamefont{Redi}},
  \bibinfo{author}{\bibfnamefont{F.}~\bibnamefont{Riva}},
  \bibinfo{author}{\bibfnamefont{A.}~\bibnamefont{Strumia}}, \bibnamefont{and}
  \bibinfo{author}{\bibfnamefont{R.}~\bibnamefont{Torre}}
  (\bibinfo{year}{2015}), \eprint{arXiv:1512.04933}.

\bibitem[{\citenamefont{Di~Chiara et~al.}(2015)\citenamefont{Di~Chiara,
  Marzola, and Raidal}}]{DiChiara:2015vdm}
\bibinfo{author}{\bibfnamefont{S.}~\bibnamefont{Di~Chiara}},
  \bibinfo{author}{\bibfnamefont{L.}~\bibnamefont{Marzola}}, \bibnamefont{and}
  \bibinfo{author}{\bibfnamefont{M.}~\bibnamefont{Raidal}}
  (\bibinfo{year}{2015}), \eprint{arXiv:1512.04939}.

\bibitem[{\citenamefont{McDermott et~al.}(2015)\citenamefont{McDermott, Meade,
  and Ramani}}]{McDermott:2015sck}
\bibinfo{author}{\bibfnamefont{S.~D.} \bibnamefont{McDermott}},
  \bibinfo{author}{\bibfnamefont{P.}~\bibnamefont{Meade}}, \bibnamefont{and}
  \bibinfo{author}{\bibfnamefont{H.}~\bibnamefont{Ramani}}
  (\bibinfo{year}{2015}), \eprint{arXiv:1512.05326}.

\bibitem[{\citenamefont{Ellis et~al.}(2015)\citenamefont{Ellis, Ellis,
  Quevillon, Sanz, and You}}]{Ellis:2015oso}
\bibinfo{author}{\bibfnamefont{J.}~\bibnamefont{Ellis}},
  \bibinfo{author}{\bibfnamefont{S.~A.~R.} \bibnamefont{Ellis}},
  \bibinfo{author}{\bibfnamefont{J.}~\bibnamefont{Quevillon}},
  \bibinfo{author}{\bibfnamefont{V.}~\bibnamefont{Sanz}}, \bibnamefont{and}
  \bibinfo{author}{\bibfnamefont{T.}~\bibnamefont{You}} (\bibinfo{year}{2015}),
  \eprint{arXiv:1512.05327}.

\bibitem[{\citenamefont{Low et~al.}(2015)\citenamefont{Low, Tesi, and
  Wang}}]{Low:2015qep}
\bibinfo{author}{\bibfnamefont{M.}~\bibnamefont{Low}},
  \bibinfo{author}{\bibfnamefont{A.}~\bibnamefont{Tesi}}, \bibnamefont{and}
  \bibinfo{author}{\bibfnamefont{L.-T.} \bibnamefont{Wang}}
  (\bibinfo{year}{2015}), \eprint{arXiv:1512.05328}.

\bibitem[{\citenamefont{Bellazzini et~al.}(2015)\citenamefont{Bellazzini,
  Franceschini, Sala, and Serra}}]{Bellazzini:2015nxw}
\bibinfo{author}{\bibfnamefont{B.}~\bibnamefont{Bellazzini}},
  \bibinfo{author}{\bibfnamefont{R.}~\bibnamefont{Franceschini}},
  \bibinfo{author}{\bibfnamefont{F.}~\bibnamefont{Sala}}, \bibnamefont{and}
  \bibinfo{author}{\bibfnamefont{J.}~\bibnamefont{Serra}}
  (\bibinfo{year}{2015}), \eprint{arXiv:1512.05330}.

\bibitem[{\citenamefont{Gupta et~al.}(2015)\citenamefont{Gupta, Jäger, Kats,
  Perez, and Stamou}}]{Gupta:2015zzs}
\bibinfo{author}{\bibfnamefont{R.~S.} \bibnamefont{Gupta}},
  \bibinfo{author}{\bibfnamefont{S.}~\bibnamefont{Jäger}},
  \bibinfo{author}{\bibfnamefont{Y.}~\bibnamefont{Kats}},
  \bibinfo{author}{\bibfnamefont{G.}~\bibnamefont{Perez}}, \bibnamefont{and}
  \bibinfo{author}{\bibfnamefont{E.}~\bibnamefont{Stamou}}
  (\bibinfo{year}{2015}), \eprint{arXiv:1512.05332}.

\bibitem[{\citenamefont{Petersson and Torre}(2015)}]{Petersson:2015mkr}
\bibinfo{author}{\bibfnamefont{C.}~\bibnamefont{Petersson}} \bibnamefont{and}
  \bibinfo{author}{\bibfnamefont{R.}~\bibnamefont{Torre}}
  (\bibinfo{year}{2015}), \eprint{arXiv:1512.05333}.

\bibitem[{\citenamefont{Molinaro et~al.}(2015)\citenamefont{Molinaro, Sannino,
  and Vignaroli}}]{Molinaro:2015cwg}
\bibinfo{author}{\bibfnamefont{E.}~\bibnamefont{Molinaro}},
  \bibinfo{author}{\bibfnamefont{F.}~\bibnamefont{Sannino}}, \bibnamefont{and}
  \bibinfo{author}{\bibfnamefont{N.}~\bibnamefont{Vignaroli}}
  (\bibinfo{year}{2015}), \eprint{arXiv:1512.05334}.

\bibitem[{\citenamefont{Costa et~al.}(2015)\citenamefont{Costa, Mühlleitner,
  Sampaio, and Santos}}]{Costa:2015llh}
\bibinfo{author}{\bibfnamefont{R.}~\bibnamefont{Costa}},
  \bibinfo{author}{\bibfnamefont{M.}~\bibnamefont{Mühlleitner}},
  \bibinfo{author}{\bibfnamefont{M.~O.~P.} \bibnamefont{Sampaio}},
  \bibnamefont{and} \bibinfo{author}{\bibfnamefont{R.}~\bibnamefont{Santos}}
  (\bibinfo{year}{2015}), \eprint{arXiv:1512.05355}.

\bibitem[{\citenamefont{Dutta et~al.}(2015)\citenamefont{Dutta, Gao, Ghosh,
  Gogoladze, and Li}}]{Dutta:2015wqh}
\bibinfo{author}{\bibfnamefont{B.}~\bibnamefont{Dutta}},
  \bibinfo{author}{\bibfnamefont{Y.}~\bibnamefont{Gao}},
  \bibinfo{author}{\bibfnamefont{T.}~\bibnamefont{Ghosh}},
  \bibinfo{author}{\bibfnamefont{I.}~\bibnamefont{Gogoladze}},
  \bibnamefont{and} \bibinfo{author}{\bibfnamefont{T.}~\bibnamefont{Li}}
  (\bibinfo{year}{2015}), \eprint{arXiv:1512.05439}.

\bibitem[{\citenamefont{Cao et~al.}(2015{\natexlab{a}})\citenamefont{Cao, Liu,
  Xie, Yan, and Zhang}}]{Cao:2015pto}
\bibinfo{author}{\bibfnamefont{Q.-H.} \bibnamefont{Cao}},
  \bibinfo{author}{\bibfnamefont{Y.}~\bibnamefont{Liu}},
  \bibinfo{author}{\bibfnamefont{K.-P.} \bibnamefont{Xie}},
  \bibinfo{author}{\bibfnamefont{B.}~\bibnamefont{Yan}}, \bibnamefont{and}
  \bibinfo{author}{\bibfnamefont{D.-M.} \bibnamefont{Zhang}}
  (\bibinfo{year}{2015}{\natexlab{a}}), \eprint{arXiv:1512.05542}.

\bibitem[{\citenamefont{Yamatsu}(2015)}]{Yamatsu:2015oit}
\bibinfo{author}{\bibfnamefont{N.}~\bibnamefont{Yamatsu}}
  (\bibinfo{year}{2015}), \eprint{arXiv:1512.05559}.

\bibitem[{\citenamefont{Matsuzaki and Yamawaki}(2015)}]{Matsuzaki:2015che}
\bibinfo{author}{\bibfnamefont{S.}~\bibnamefont{Matsuzaki}} \bibnamefont{and}
  \bibinfo{author}{\bibfnamefont{K.}~\bibnamefont{Yamawaki}}
  (\bibinfo{year}{2015}), \eprint{arXiv:1512.05564}.

\bibitem[{\citenamefont{Kobakhidze et~al.}(2015)\citenamefont{Kobakhidze, Wang,
  Wu, Yang, and Zhang}}]{Kobakhidze:2015ldh}
\bibinfo{author}{\bibfnamefont{A.}~\bibnamefont{Kobakhidze}},
  \bibinfo{author}{\bibfnamefont{F.}~\bibnamefont{Wang}},
  \bibinfo{author}{\bibfnamefont{L.}~\bibnamefont{Wu}},
  \bibinfo{author}{\bibfnamefont{J.~M.} \bibnamefont{Yang}}, \bibnamefont{and}
  \bibinfo{author}{\bibfnamefont{M.}~\bibnamefont{Zhang}}
  (\bibinfo{year}{2015}), \eprint{arXiv:1512.05585}.

\bibitem[{\citenamefont{Martinez et~al.}(2015)\citenamefont{Martinez, Ochoa,
  and Sierra}}]{Martinez:2015kmn}
\bibinfo{author}{\bibfnamefont{R.}~\bibnamefont{Martinez}},
  \bibinfo{author}{\bibfnamefont{F.}~\bibnamefont{Ochoa}}, \bibnamefont{and}
  \bibinfo{author}{\bibfnamefont{C.~F.} \bibnamefont{Sierra}}
  (\bibinfo{year}{2015}), \eprint{arXiv:1512.05617}.

\bibitem[{\citenamefont{Cox et~al.}(2015)\citenamefont{Cox, Medina, Ray, and
  Spray}}]{Cox:2015ckc}
\bibinfo{author}{\bibfnamefont{P.}~\bibnamefont{Cox}},
  \bibinfo{author}{\bibfnamefont{A.~D.} \bibnamefont{Medina}},
  \bibinfo{author}{\bibfnamefont{T.~S.} \bibnamefont{Ray}}, \bibnamefont{and}
  \bibinfo{author}{\bibfnamefont{A.}~\bibnamefont{Spray}}
  (\bibinfo{year}{2015}), \eprint{arXiv:1512.05618}.

\bibitem[{\citenamefont{Becirevic et~al.}(2015)\citenamefont{Becirevic,
  Bertuzzo, Sumensari, and Funchal}}]{Becirevic:2015fmu}
\bibinfo{author}{\bibfnamefont{D.}~\bibnamefont{Becirevic}},
  \bibinfo{author}{\bibfnamefont{E.}~\bibnamefont{Bertuzzo}},
  \bibinfo{author}{\bibfnamefont{O.}~\bibnamefont{Sumensari}},
  \bibnamefont{and} \bibinfo{author}{\bibfnamefont{R.~Z.}
  \bibnamefont{Funchal}} (\bibinfo{year}{2015}), \eprint{arXiv:1512.05623}.

\bibitem[{\citenamefont{No et~al.}(2015)\citenamefont{No, Sanz, and
  Setford}}]{No:2015bsn}
\bibinfo{author}{\bibfnamefont{J.~M.} \bibnamefont{No}},
  \bibinfo{author}{\bibfnamefont{V.}~\bibnamefont{Sanz}}, \bibnamefont{and}
  \bibinfo{author}{\bibfnamefont{J.}~\bibnamefont{Setford}}
  (\bibinfo{year}{2015}), \eprint{arXiv:1512.05700}.

\bibitem[{\citenamefont{Demidov and Gorbunov}(2015)}]{Demidov:2015zqn}
\bibinfo{author}{\bibfnamefont{S.~V.} \bibnamefont{Demidov}} \bibnamefont{and}
  \bibinfo{author}{\bibfnamefont{D.~S.} \bibnamefont{Gorbunov}}
  (\bibinfo{year}{2015}), \eprint{arXiv:1512.05723}.

\bibitem[{\citenamefont{Gopalakrishna et~al.}(2015)\citenamefont{Gopalakrishna,
  Mukherjee, and Sadhukhan}}]{Gopalakrishna:2015dkt}
\bibinfo{author}{\bibfnamefont{S.}~\bibnamefont{Gopalakrishna}},
  \bibinfo{author}{\bibfnamefont{T.~S.} \bibnamefont{Mukherjee}},
  \bibnamefont{and} \bibinfo{author}{\bibfnamefont{S.}~\bibnamefont{Sadhukhan}}
  (\bibinfo{year}{2015}), \eprint{arXiv:1512.05731}.

\bibitem[{\citenamefont{Chao et~al.}(2015)\citenamefont{Chao, Huo, and
  Yu}}]{Chao:2015ttq}
\bibinfo{author}{\bibfnamefont{W.}~\bibnamefont{Chao}},
  \bibinfo{author}{\bibfnamefont{R.}~\bibnamefont{Huo}}, \bibnamefont{and}
  \bibinfo{author}{\bibfnamefont{J.-H.} \bibnamefont{Yu}}
  (\bibinfo{year}{2015}), \eprint{arXiv:1512.05738}.

\bibitem[{\citenamefont{Fichet et~al.}(2015)\citenamefont{Fichet, von
  Gersdorff, and Royon}}]{Fichet:2015vvy}
\bibinfo{author}{\bibfnamefont{S.}~\bibnamefont{Fichet}},
  \bibinfo{author}{\bibfnamefont{G.}~\bibnamefont{von Gersdorff}},
  \bibnamefont{and} \bibinfo{author}{\bibfnamefont{C.}~\bibnamefont{Royon}}
  (\bibinfo{year}{2015}), \eprint{arXiv:1512.05751}.

\bibitem[{\citenamefont{Bian et~al.}(2015)\citenamefont{Bian, Chen, Liu, and
  Shu}}]{Bian:2015kjt}
\bibinfo{author}{\bibfnamefont{L.}~\bibnamefont{Bian}},
  \bibinfo{author}{\bibfnamefont{N.}~\bibnamefont{Chen}},
  \bibinfo{author}{\bibfnamefont{D.}~\bibnamefont{Liu}}, \bibnamefont{and}
  \bibinfo{author}{\bibfnamefont{J.}~\bibnamefont{Shu}} (\bibinfo{year}{2015}),
  \eprint{arXiv:1512.05759}.

\bibitem[{\citenamefont{Chakrabortty et~al.}(2015)\citenamefont{Chakrabortty,
  Choudhury, Ghosh, Mondal, and Srivastava}}]{Chakrabortty:2015hff}
\bibinfo{author}{\bibfnamefont{J.}~\bibnamefont{Chakrabortty}},
  \bibinfo{author}{\bibfnamefont{A.}~\bibnamefont{Choudhury}},
  \bibinfo{author}{\bibfnamefont{P.}~\bibnamefont{Ghosh}},
  \bibinfo{author}{\bibfnamefont{S.}~\bibnamefont{Mondal}}, \bibnamefont{and}
  \bibinfo{author}{\bibfnamefont{T.}~\bibnamefont{Srivastava}}
  (\bibinfo{year}{2015}), \eprint{arXiv:1512.05767}.

\bibitem[{\citenamefont{Ahmed et~al.}(2015)\citenamefont{Ahmed, Dillon,
  Grzadkowski, Gunion, and Jiang}}]{Ahmed:2015uqt}
\bibinfo{author}{\bibfnamefont{A.}~\bibnamefont{Ahmed}},
  \bibinfo{author}{\bibfnamefont{B.~M.} \bibnamefont{Dillon}},
  \bibinfo{author}{\bibfnamefont{B.}~\bibnamefont{Grzadkowski}},
  \bibinfo{author}{\bibfnamefont{J.~F.} \bibnamefont{Gunion}},
  \bibnamefont{and} \bibinfo{author}{\bibfnamefont{Y.}~\bibnamefont{Jiang}}
  (\bibinfo{year}{2015}), \eprint{arXiv:1512.05771}.

\bibitem[{\citenamefont{Agrawal et~al.}(2015)\citenamefont{Agrawal, Fan,
  Heidenreich, Reece, and Strassler}}]{Agrawal:2015dbf}
\bibinfo{author}{\bibfnamefont{P.}~\bibnamefont{Agrawal}},
  \bibinfo{author}{\bibfnamefont{J.}~\bibnamefont{Fan}},
  \bibinfo{author}{\bibfnamefont{B.}~\bibnamefont{Heidenreich}},
  \bibinfo{author}{\bibfnamefont{M.}~\bibnamefont{Reece}}, \bibnamefont{and}
  \bibinfo{author}{\bibfnamefont{M.}~\bibnamefont{Strassler}}
  (\bibinfo{year}{2015}), \eprint{arXiv:1512.05775}.

\bibitem[{\citenamefont{Csaki et~al.}(2015)\citenamefont{Csaki, Hubisz, and
  Terning}}]{Csaki:2015vek}
\bibinfo{author}{\bibfnamefont{C.}~\bibnamefont{Csaki}},
  \bibinfo{author}{\bibfnamefont{J.}~\bibnamefont{Hubisz}}, \bibnamefont{and}
  \bibinfo{author}{\bibfnamefont{J.}~\bibnamefont{Terning}}
  (\bibinfo{year}{2015}), \eprint{arXiv:1512.05776}.

\bibitem[{\citenamefont{Falkowski et~al.}(2015)\citenamefont{Falkowski, Slone,
  and Volansky}}]{Falkowski:2015swt}
\bibinfo{author}{\bibfnamefont{A.}~\bibnamefont{Falkowski}},
  \bibinfo{author}{\bibfnamefont{O.}~\bibnamefont{Slone}}, \bibnamefont{and}
  \bibinfo{author}{\bibfnamefont{T.}~\bibnamefont{Volansky}}
  (\bibinfo{year}{2015}), \eprint{arXiv:1512.05777}.

\bibitem[{\citenamefont{Aloni et~al.}(2015)\citenamefont{Aloni, Blum, Dery,
  Efrati, and Nir}}]{Aloni:2015mxa}
\bibinfo{author}{\bibfnamefont{D.}~\bibnamefont{Aloni}},
  \bibinfo{author}{\bibfnamefont{K.}~\bibnamefont{Blum}},
  \bibinfo{author}{\bibfnamefont{A.}~\bibnamefont{Dery}},
  \bibinfo{author}{\bibfnamefont{A.}~\bibnamefont{Efrati}}, \bibnamefont{and}
  \bibinfo{author}{\bibfnamefont{Y.}~\bibnamefont{Nir}} (\bibinfo{year}{2015}),
  \eprint{arXiv:1512.05778}.

\bibitem[{\citenamefont{Bai et~al.}(2015)\citenamefont{Bai, Berger, and
  Lu}}]{Bai:2015nbs}
\bibinfo{author}{\bibfnamefont{Y.}~\bibnamefont{Bai}},
  \bibinfo{author}{\bibfnamefont{J.}~\bibnamefont{Berger}}, \bibnamefont{and}
  \bibinfo{author}{\bibfnamefont{R.}~\bibnamefont{Lu}} (\bibinfo{year}{2015}),
  \eprint{arXiv:1512.05779}.

\bibitem[{\citenamefont{Gabrielli et~al.}(2015)\citenamefont{Gabrielli,
  Kannike, Mele, Raidal, Spethmann, and Veermäe}}]{Gabrielli:2015dhk}
\bibinfo{author}{\bibfnamefont{E.}~\bibnamefont{Gabrielli}},
  \bibinfo{author}{\bibfnamefont{K.}~\bibnamefont{Kannike}},
  \bibinfo{author}{\bibfnamefont{B.}~\bibnamefont{Mele}},
  \bibinfo{author}{\bibfnamefont{M.}~\bibnamefont{Raidal}},
  \bibinfo{author}{\bibfnamefont{C.}~\bibnamefont{Spethmann}},
  \bibnamefont{and} \bibinfo{author}{\bibfnamefont{H.}~\bibnamefont{Veermäe}}
  (\bibinfo{year}{2015}), \eprint{arXiv:1512.05961}.

\bibitem[{\citenamefont{Benbrik et~al.}(2015)\citenamefont{Benbrik, Chen, and
  Nomura}}]{Benbrik:2015fyz}
\bibinfo{author}{\bibfnamefont{R.}~\bibnamefont{Benbrik}},
  \bibinfo{author}{\bibfnamefont{C.-H.} \bibnamefont{Chen}}, \bibnamefont{and}
  \bibinfo{author}{\bibfnamefont{T.}~\bibnamefont{Nomura}}
  (\bibinfo{year}{2015}), \eprint{arXiv:1512.06028}.

\bibitem[{\citenamefont{Kim et~al.}(2015{\natexlab{a}})\citenamefont{Kim,
  Reuter, Rolbiecki, and de~Austri}}]{Kim:2015ron}
\bibinfo{author}{\bibfnamefont{J.~S.} \bibnamefont{Kim}},
  \bibinfo{author}{\bibfnamefont{J.}~\bibnamefont{Reuter}},
  \bibinfo{author}{\bibfnamefont{K.}~\bibnamefont{Rolbiecki}},
  \bibnamefont{and} \bibinfo{author}{\bibfnamefont{R.~R.}
  \bibnamefont{de~Austri}} (\bibinfo{year}{2015}{\natexlab{a}}),
  \eprint{arXiv:1512.06083}.

\bibitem[{\citenamefont{Alves et~al.}(2015)\citenamefont{Alves, Dias, and
  Sinha}}]{Alves:2015jgx}
\bibinfo{author}{\bibfnamefont{A.}~\bibnamefont{Alves}},
  \bibinfo{author}{\bibfnamefont{A.~G.} \bibnamefont{Dias}}, \bibnamefont{and}
  \bibinfo{author}{\bibfnamefont{K.}~\bibnamefont{Sinha}}
  (\bibinfo{year}{2015}), \eprint{arXiv:1512.06091}.

\bibitem[{\citenamefont{Megias et~al.}(2015)\citenamefont{Megias, Pujolas, and
  Quiros}}]{Megias:2015ory}
\bibinfo{author}{\bibfnamefont{E.}~\bibnamefont{Megias}},
  \bibinfo{author}{\bibfnamefont{O.}~\bibnamefont{Pujolas}}, \bibnamefont{and}
  \bibinfo{author}{\bibfnamefont{M.}~\bibnamefont{Quiros}}
  (\bibinfo{year}{2015}), \eprint{arXiv:1512.06106}.

\bibitem[{\citenamefont{Carpenter et~al.}(2015)\citenamefont{Carpenter,
  Colburn, and Goodman}}]{Carpenter:2015ucu}
\bibinfo{author}{\bibfnamefont{L.~M.} \bibnamefont{Carpenter}},
  \bibinfo{author}{\bibfnamefont{R.}~\bibnamefont{Colburn}}, \bibnamefont{and}
  \bibinfo{author}{\bibfnamefont{J.}~\bibnamefont{Goodman}}
  (\bibinfo{year}{2015}), \eprint{arXiv:1512.06107}.

\bibitem[{\citenamefont{Chao}(2015)}]{Chao:2015nsm}
\bibinfo{author}{\bibfnamefont{W.}~\bibnamefont{Chao}} (\bibinfo{year}{2015}),
  \eprint{arXiv:1512.06297}.

\bibitem[{\citenamefont{Arun and Saha}(2015)}]{Arun:2015ubr}
\bibinfo{author}{\bibfnamefont{M.~T.} \bibnamefont{Arun}} \bibnamefont{and}
  \bibinfo{author}{\bibfnamefont{P.}~\bibnamefont{Saha}}
  (\bibinfo{year}{2015}), \eprint{arXiv:1512.06335}.

\bibitem[{\citenamefont{Han et~al.}(2015{\natexlab{a}})\citenamefont{Han, Lee,
  Park, and Sanz}}]{Han:2015cty}
\bibinfo{author}{\bibfnamefont{C.}~\bibnamefont{Han}},
  \bibinfo{author}{\bibfnamefont{H.~M.} \bibnamefont{Lee}},
  \bibinfo{author}{\bibfnamefont{M.}~\bibnamefont{Park}}, \bibnamefont{and}
  \bibinfo{author}{\bibfnamefont{V.}~\bibnamefont{Sanz}}
  (\bibinfo{year}{2015}{\natexlab{a}}), \eprint{arXiv:1512.06376}.

\bibitem[{\citenamefont{Chang}(2015)}]{Chang:2015bzc}
\bibinfo{author}{\bibfnamefont{S.}~\bibnamefont{Chang}} (\bibinfo{year}{2015}),
  \eprint{arXiv:1512.06426}.

\bibitem[{\citenamefont{Chakraborty and Kundu}(2015)}]{Chakraborty:2015jvs}
\bibinfo{author}{\bibfnamefont{I.}~\bibnamefont{Chakraborty}} \bibnamefont{and}
  \bibinfo{author}{\bibfnamefont{A.}~\bibnamefont{Kundu}}
  (\bibinfo{year}{2015}), \eprint{arXiv:1512.06508}.

\bibitem[{\citenamefont{Ding et~al.}(2015)\citenamefont{Ding, Huang, Li, and
  Zhu}}]{Ding:2015rxx}
\bibinfo{author}{\bibfnamefont{R.}~\bibnamefont{Ding}},
  \bibinfo{author}{\bibfnamefont{L.}~\bibnamefont{Huang}},
  \bibinfo{author}{\bibfnamefont{T.}~\bibnamefont{Li}}, \bibnamefont{and}
  \bibinfo{author}{\bibfnamefont{B.}~\bibnamefont{Zhu}} (\bibinfo{year}{2015}),
  \eprint{arXiv:1512.06560}.

\bibitem[{\citenamefont{Han et~al.}(2015{\natexlab{b}})\citenamefont{Han, Wang,
  and Zheng}}]{Han:2015dlp}
\bibinfo{author}{\bibfnamefont{H.}~\bibnamefont{Han}},
  \bibinfo{author}{\bibfnamefont{S.}~\bibnamefont{Wang}}, \bibnamefont{and}
  \bibinfo{author}{\bibfnamefont{S.}~\bibnamefont{Zheng}}
  (\bibinfo{year}{2015}{\natexlab{b}}), \eprint{arXiv:1512.06562}.

\bibitem[{\citenamefont{Han and Wang}(2015)}]{Han:2015qqj}
\bibinfo{author}{\bibfnamefont{X.-F.} \bibnamefont{Han}} \bibnamefont{and}
  \bibinfo{author}{\bibfnamefont{L.}~\bibnamefont{Wang}}
  (\bibinfo{year}{2015}), \eprint{arXiv:1512.06587}.

\bibitem[{\citenamefont{Luo et~al.}(2015)\citenamefont{Luo, Wang, Xu, Zhang,
  and Zhu}}]{Luo:2015yio}
\bibinfo{author}{\bibfnamefont{M.-x.} \bibnamefont{Luo}},
  \bibinfo{author}{\bibfnamefont{K.}~\bibnamefont{Wang}},
  \bibinfo{author}{\bibfnamefont{T.}~\bibnamefont{Xu}},
  \bibinfo{author}{\bibfnamefont{L.}~\bibnamefont{Zhang}}, \bibnamefont{and}
  \bibinfo{author}{\bibfnamefont{G.}~\bibnamefont{Zhu}} (\bibinfo{year}{2015}),
  \eprint{arXiv:1512.06670}.

\bibitem[{\citenamefont{Chang et~al.}(2015)\citenamefont{Chang, Cheung, and
  Lu}}]{Chang:2015sdy}
\bibinfo{author}{\bibfnamefont{J.}~\bibnamefont{Chang}},
  \bibinfo{author}{\bibfnamefont{K.}~\bibnamefont{Cheung}}, \bibnamefont{and}
  \bibinfo{author}{\bibfnamefont{C.-T.} \bibnamefont{Lu}}
  (\bibinfo{year}{2015}), \eprint{arXiv:1512.06671}.

\bibitem[{\citenamefont{Bardhan et~al.}(2015)\citenamefont{Bardhan, Bhatia,
  Chakraborty, Maitra, Raychaudhuri, and Samui}}]{Bardhan:2015hcr}
\bibinfo{author}{\bibfnamefont{D.}~\bibnamefont{Bardhan}},
  \bibinfo{author}{\bibfnamefont{D.}~\bibnamefont{Bhatia}},
  \bibinfo{author}{\bibfnamefont{A.}~\bibnamefont{Chakraborty}},
  \bibinfo{author}{\bibfnamefont{U.}~\bibnamefont{Maitra}},
  \bibinfo{author}{\bibfnamefont{S.}~\bibnamefont{Raychaudhuri}},
  \bibnamefont{and} \bibinfo{author}{\bibfnamefont{T.}~\bibnamefont{Samui}}
  (\bibinfo{year}{2015}), \eprint{arXiv:1512.06674}.

\bibitem[{\citenamefont{Feng et~al.}(2015)\citenamefont{Feng, Li, Zhang, and
  Zhao}}]{Feng:2015wil}
\bibinfo{author}{\bibfnamefont{T.-F.} \bibnamefont{Feng}},
  \bibinfo{author}{\bibfnamefont{X.-Q.} \bibnamefont{Li}},
  \bibinfo{author}{\bibfnamefont{H.-B.} \bibnamefont{Zhang}}, \bibnamefont{and}
  \bibinfo{author}{\bibfnamefont{S.-M.} \bibnamefont{Zhao}}
  (\bibinfo{year}{2015}), \eprint{arXiv:1512.06696}.

\bibitem[{\citenamefont{Wang et~al.}(2015)\citenamefont{Wang, Wu, Yang, and
  Zhang}}]{Wang:2015kuj}
\bibinfo{author}{\bibfnamefont{F.}~\bibnamefont{Wang}},
  \bibinfo{author}{\bibfnamefont{L.}~\bibnamefont{Wu}},
  \bibinfo{author}{\bibfnamefont{J.~M.} \bibnamefont{Yang}}, \bibnamefont{and}
  \bibinfo{author}{\bibfnamefont{M.}~\bibnamefont{Zhang}}
  (\bibinfo{year}{2015}), \eprint{arXiv:1512.06715}.

\bibitem[{\citenamefont{Cao et~al.}(2015{\natexlab{b}})\citenamefont{Cao, Han,
  Shang, Su, Yang, and Zhang}}]{Cao:2015twy}
\bibinfo{author}{\bibfnamefont{J.}~\bibnamefont{Cao}},
  \bibinfo{author}{\bibfnamefont{C.}~\bibnamefont{Han}},
  \bibinfo{author}{\bibfnamefont{L.}~\bibnamefont{Shang}},
  \bibinfo{author}{\bibfnamefont{W.}~\bibnamefont{Su}},
  \bibinfo{author}{\bibfnamefont{J.~M.} \bibnamefont{Yang}}, \bibnamefont{and}
  \bibinfo{author}{\bibfnamefont{Y.}~\bibnamefont{Zhang}}
  (\bibinfo{year}{2015}{\natexlab{b}}), \eprint{arXiv:1512.06728}.

\bibitem[{\citenamefont{Huang et~al.}(2015{\natexlab{a}})\citenamefont{Huang,
  Li, Liu, and Wang}}]{Huang:2015evq}
\bibinfo{author}{\bibfnamefont{F.~P.} \bibnamefont{Huang}},
  \bibinfo{author}{\bibfnamefont{C.~S.} \bibnamefont{Li}},
  \bibinfo{author}{\bibfnamefont{Z.~L.} \bibnamefont{Liu}}, \bibnamefont{and}
  \bibinfo{author}{\bibfnamefont{Y.}~\bibnamefont{Wang}}
  (\bibinfo{year}{2015}{\natexlab{a}}), \eprint{arXiv:1512.06732}.

\bibitem[{\citenamefont{Liao and Zheng}(2015)}]{Liao:2015tow}
\bibinfo{author}{\bibfnamefont{W.}~\bibnamefont{Liao}} \bibnamefont{and}
  \bibinfo{author}{\bibfnamefont{H.-q.} \bibnamefont{Zheng}}
  (\bibinfo{year}{2015}), \eprint{arXiv:1512.06741}.

\bibitem[{\citenamefont{Heckman}(2015)}]{Heckman:2015kqk}
\bibinfo{author}{\bibfnamefont{J.~J.} \bibnamefont{Heckman}}
  (\bibinfo{year}{2015}), \eprint{arXiv:1512.06773}.

\bibitem[{\citenamefont{Dhuria and Goswami}(2015)}]{Dhuria:2015ufo}
\bibinfo{author}{\bibfnamefont{M.}~\bibnamefont{Dhuria}} \bibnamefont{and}
  \bibinfo{author}{\bibfnamefont{G.}~\bibnamefont{Goswami}}
  (\bibinfo{year}{2015}), \eprint{arXiv:1512.06782}.

\bibitem[{\citenamefont{Bi et~al.}(2015)\citenamefont{Bi, Xiang, Yin, and
  Yu}}]{Bi:2015uqd}
\bibinfo{author}{\bibfnamefont{X.-J.} \bibnamefont{Bi}},
  \bibinfo{author}{\bibfnamefont{Q.-F.} \bibnamefont{Xiang}},
  \bibinfo{author}{\bibfnamefont{P.-F.} \bibnamefont{Yin}}, \bibnamefont{and}
  \bibinfo{author}{\bibfnamefont{Z.-H.} \bibnamefont{Yu}}
  (\bibinfo{year}{2015}), \eprint{arXiv:1512.06787}.

\bibitem[{\citenamefont{Kim et~al.}(2015{\natexlab{b}})\citenamefont{Kim,
  Rolbiecki, and de~Austri}}]{Kim:2015ksf}
\bibinfo{author}{\bibfnamefont{J.~S.} \bibnamefont{Kim}},
  \bibinfo{author}{\bibfnamefont{K.}~\bibnamefont{Rolbiecki}},
  \bibnamefont{and} \bibinfo{author}{\bibfnamefont{R.~R.}
  \bibnamefont{de~Austri}} (\bibinfo{year}{2015}{\natexlab{b}}),
  \eprint{arXiv:1512.06797}.

\bibitem[{\citenamefont{Berthier et~al.}(2015)\citenamefont{Berthier, Cline,
  Shepherd, and Trott}}]{Berthier:2015vbb}
\bibinfo{author}{\bibfnamefont{L.}~\bibnamefont{Berthier}},
  \bibinfo{author}{\bibfnamefont{J.~M.} \bibnamefont{Cline}},
  \bibinfo{author}{\bibfnamefont{W.}~\bibnamefont{Shepherd}}, \bibnamefont{and}
  \bibinfo{author}{\bibfnamefont{M.}~\bibnamefont{Trott}}
  (\bibinfo{year}{2015}), \eprint{arXiv:1512.06799}.

\bibitem[{\citenamefont{Cho et~al.}(2015)\citenamefont{Cho, Kim, Kong, Lim,
  Matchev, Park, and Park}}]{Cho:2015nxy}
\bibinfo{author}{\bibfnamefont{W.~S.} \bibnamefont{Cho}},
  \bibinfo{author}{\bibfnamefont{D.}~\bibnamefont{Kim}},
  \bibinfo{author}{\bibfnamefont{K.}~\bibnamefont{Kong}},
  \bibinfo{author}{\bibfnamefont{S.~H.} \bibnamefont{Lim}},
  \bibinfo{author}{\bibfnamefont{K.~T.} \bibnamefont{Matchev}},
  \bibinfo{author}{\bibfnamefont{J.-C.} \bibnamefont{Park}}, \bibnamefont{and}
  \bibinfo{author}{\bibfnamefont{M.}~\bibnamefont{Park}}
  (\bibinfo{year}{2015}), \eprint{arXiv:1512.06824}.

\bibitem[{\citenamefont{Cline and Liu}(2015)}]{Cline:2015msi}
\bibinfo{author}{\bibfnamefont{J.~M.} \bibnamefont{Cline}} \bibnamefont{and}
  \bibinfo{author}{\bibfnamefont{Z.}~\bibnamefont{Liu}} (\bibinfo{year}{2015}),
  \eprint{arXiv:1512.06827}.

\bibitem[{\citenamefont{Chala et~al.}(2015)\citenamefont{Chala, Duerr,
  Kahlhoefer, and Schmidt-Hoberg}}]{Chala:2015cev}
\bibinfo{author}{\bibfnamefont{M.}~\bibnamefont{Chala}},
  \bibinfo{author}{\bibfnamefont{M.}~\bibnamefont{Duerr}},
  \bibinfo{author}{\bibfnamefont{F.}~\bibnamefont{Kahlhoefer}},
  \bibnamefont{and}
  \bibinfo{author}{\bibfnamefont{K.}~\bibnamefont{Schmidt-Hoberg}}
  (\bibinfo{year}{2015}), \eprint{arXiv:1512.06833}.

\bibitem[{\citenamefont{Barducci et~al.}(2015)\citenamefont{Barducci, Goudelis,
  Kulkarni, and Sengupta}}]{Barducci:2015gtd}
\bibinfo{author}{\bibfnamefont{D.}~\bibnamefont{Barducci}},
  \bibinfo{author}{\bibfnamefont{A.}~\bibnamefont{Goudelis}},
  \bibinfo{author}{\bibfnamefont{S.}~\bibnamefont{Kulkarni}}, \bibnamefont{and}
  \bibinfo{author}{\bibfnamefont{D.}~\bibnamefont{Sengupta}}
  (\bibinfo{year}{2015}), \eprint{arXiv:1512.06842}.

\bibitem[{\citenamefont{Boucenna et~al.}(2015)\citenamefont{Boucenna, Morisi,
  and Vicente}}]{Boucenna:2015pav}
\bibinfo{author}{\bibfnamefont{S.~M.} \bibnamefont{Boucenna}},
  \bibinfo{author}{\bibfnamefont{S.}~\bibnamefont{Morisi}}, \bibnamefont{and}
  \bibinfo{author}{\bibfnamefont{A.}~\bibnamefont{Vicente}}
  (\bibinfo{year}{2015}), \eprint{arXiv:1512.06878}.

\bibitem[{\citenamefont{Murphy}(2015)}]{Murphy:2015kag}
\bibinfo{author}{\bibfnamefont{C.~W.} \bibnamefont{Murphy}}
  (\bibinfo{year}{2015}), \eprint{arXiv:1512.06976}.

\bibitem[{\citenamefont{Hernández and Nisandzic}(2015)}]{Hernandez:2015ywg}
\bibinfo{author}{\bibfnamefont{A.~E.~C.} \bibnamefont{Hernández}}
  \bibnamefont{and} \bibinfo{author}{\bibfnamefont{I.}~\bibnamefont{Nisandzic}}
  (\bibinfo{year}{2015}), \eprint{arXiv:1512.07165}.

\bibitem[{\citenamefont{Dey et~al.}(2015)\citenamefont{Dey, Mohanty, and
  Tomar}}]{Dey:2015bur}
\bibinfo{author}{\bibfnamefont{U.~K.} \bibnamefont{Dey}},
  \bibinfo{author}{\bibfnamefont{S.}~\bibnamefont{Mohanty}}, \bibnamefont{and}
  \bibinfo{author}{\bibfnamefont{G.}~\bibnamefont{Tomar}}
  (\bibinfo{year}{2015}), \eprint{arXiv:1512.07212}.

\bibitem[{\citenamefont{Pelaggi et~al.}(2015)\citenamefont{Pelaggi, Strumia,
  and Vigiani}}]{Pelaggi:2015knk}
\bibinfo{author}{\bibfnamefont{G.~M.} \bibnamefont{Pelaggi}},
  \bibinfo{author}{\bibfnamefont{A.}~\bibnamefont{Strumia}}, \bibnamefont{and}
  \bibinfo{author}{\bibfnamefont{E.}~\bibnamefont{Vigiani}}
  (\bibinfo{year}{2015}), \eprint{arXiv:1512.07225}.

\bibitem[{\citenamefont{de~Blas et~al.}(2015)\citenamefont{de~Blas, Santiago,
  and Vega-Morales}}]{deBlas:2015hlv}
\bibinfo{author}{\bibfnamefont{J.}~\bibnamefont{de~Blas}},
  \bibinfo{author}{\bibfnamefont{J.}~\bibnamefont{Santiago}}, \bibnamefont{and}
  \bibinfo{author}{\bibfnamefont{R.}~\bibnamefont{Vega-Morales}}
  (\bibinfo{year}{2015}), \eprint{arXiv:1512.07229}.

\bibitem[{\citenamefont{Belyaev et~al.}(2015)\citenamefont{Belyaev,
  Cacciapaglia, Cai, Flacke, Parolini, and Serôdio}}]{Belyaev:2015hgo}
\bibinfo{author}{\bibfnamefont{A.}~\bibnamefont{Belyaev}},
  \bibinfo{author}{\bibfnamefont{G.}~\bibnamefont{Cacciapaglia}},
  \bibinfo{author}{\bibfnamefont{H.}~\bibnamefont{Cai}},
  \bibinfo{author}{\bibfnamefont{T.}~\bibnamefont{Flacke}},
  \bibinfo{author}{\bibfnamefont{A.}~\bibnamefont{Parolini}}, \bibnamefont{and}
  \bibinfo{author}{\bibfnamefont{H.}~\bibnamefont{Serôdio}}
  (\bibinfo{year}{2015}), \eprint{arXiv:1512.07242}.

\bibitem[{\citenamefont{Dev and Teresi}(2015)}]{Dev:2015isx}
\bibinfo{author}{\bibfnamefont{P.~S.~B.} \bibnamefont{Dev}} \bibnamefont{and}
  \bibinfo{author}{\bibfnamefont{D.}~\bibnamefont{Teresi}}
  (\bibinfo{year}{2015}), \eprint{arXiv:1512.07243}.

\bibitem[{\citenamefont{Huang et~al.}(2015{\natexlab{b}})\citenamefont{Huang,
  Tsai, and Yuan}}]{Huang:2015rkj}
\bibinfo{author}{\bibfnamefont{W.-C.} \bibnamefont{Huang}},
  \bibinfo{author}{\bibfnamefont{Y.-L.~S.} \bibnamefont{Tsai}},
  \bibnamefont{and} \bibinfo{author}{\bibfnamefont{T.-C.} \bibnamefont{Yuan}}
  (\bibinfo{year}{2015}{\natexlab{b}}), \eprint{arXiv:1512.07268}.

\bibitem[{\citenamefont{Moretti and Yagyu}(2015)}]{Moretti:2015pbj}
\bibinfo{author}{\bibfnamefont{S.}~\bibnamefont{Moretti}} \bibnamefont{and}
  \bibinfo{author}{\bibfnamefont{K.}~\bibnamefont{Yagyu}}
  (\bibinfo{year}{2015}), \eprint{arXiv:1512.07462}.

\bibitem[{\citenamefont{Patel and Sharma}(2015)}]{Patel:2015ulo}
\bibinfo{author}{\bibfnamefont{K.~M.} \bibnamefont{Patel}} \bibnamefont{and}
  \bibinfo{author}{\bibfnamefont{P.}~\bibnamefont{Sharma}}
  (\bibinfo{year}{2015}), \eprint{arXiv:1512.07468}.

\bibitem[{\citenamefont{Badziak}(2015)}]{Badziak:2015zez}
\bibinfo{author}{\bibfnamefont{M.}~\bibnamefont{Badziak}}
  (\bibinfo{year}{2015}), \eprint{arXiv:1512.07497}.

\bibitem[{\citenamefont{Chakraborty et~al.}(2015)\citenamefont{Chakraborty,
  Chakraborty, and Raychaudhuri}}]{Chakraborty:2015gyj}
\bibinfo{author}{\bibfnamefont{S.}~\bibnamefont{Chakraborty}},
  \bibinfo{author}{\bibfnamefont{A.}~\bibnamefont{Chakraborty}},
  \bibnamefont{and}
  \bibinfo{author}{\bibfnamefont{S.}~\bibnamefont{Raychaudhuri}}
  (\bibinfo{year}{2015}), \eprint{arXiv:1512.07527}.

\bibitem[{\citenamefont{Cao et~al.}(2015{\natexlab{c}})\citenamefont{Cao, Chen,
  and Gu}}]{Cao:2015xjz}
\bibinfo{author}{\bibfnamefont{Q.-H.} \bibnamefont{Cao}},
  \bibinfo{author}{\bibfnamefont{S.-L.} \bibnamefont{Chen}}, \bibnamefont{and}
  \bibinfo{author}{\bibfnamefont{P.-H.} \bibnamefont{Gu}}
  (\bibinfo{year}{2015}{\natexlab{c}}), \eprint{arXiv:1512.07541}.

\bibitem[{\citenamefont{Altmannshofer et~al.}(2015)\citenamefont{Altmannshofer,
  Galloway, Gori, Kagan, Martin, and Zupan}}]{Altmannshofer:2015xfo}
\bibinfo{author}{\bibfnamefont{W.}~\bibnamefont{Altmannshofer}},
  \bibinfo{author}{\bibfnamefont{J.}~\bibnamefont{Galloway}},
  \bibinfo{author}{\bibfnamefont{S.}~\bibnamefont{Gori}},
  \bibinfo{author}{\bibfnamefont{A.~L.} \bibnamefont{Kagan}},
  \bibinfo{author}{\bibfnamefont{A.}~\bibnamefont{Martin}}, \bibnamefont{and}
  \bibinfo{author}{\bibfnamefont{J.}~\bibnamefont{Zupan}}
  (\bibinfo{year}{2015}), \eprint{arXiv:1512.07616}.

\bibitem[{\citenamefont{Cveti$\text{\v{c}}$
  et~al.}(2015)\citenamefont{Cveti$\text{\v{c}}$, Halverson, and
  Langacker}}]{Cvetic:2015vit}
\bibinfo{author}{\bibfnamefont{M.}~\bibnamefont{Cveti$\text{\v{c}}$}},
  \bibinfo{author}{\bibfnamefont{J.}~\bibnamefont{Halverson}},
  \bibnamefont{and} \bibinfo{author}{\bibfnamefont{P.}~\bibnamefont{Langacker}}
  (\bibinfo{year}{2015}), \eprint{arXiv:1512.07622}.

\bibitem[{\citenamefont{Gu and Liu}(2015)}]{Gu:2015lxj}
\bibinfo{author}{\bibfnamefont{J.}~\bibnamefont{Gu}} \bibnamefont{and}
  \bibinfo{author}{\bibfnamefont{Z.}~\bibnamefont{Liu}} (\bibinfo{year}{2015}),
  \eprint{arXiv:1512.07624}.

\bibitem[{\citenamefont{Allanach et~al.}(2015)\citenamefont{Allanach, Dev,
  Renner, and Sakurai}}]{Allanach:2015ixl}
\bibinfo{author}{\bibfnamefont{B.~C.} \bibnamefont{Allanach}},
  \bibinfo{author}{\bibfnamefont{P.~S.~B.} \bibnamefont{Dev}},
  \bibinfo{author}{\bibfnamefont{S.~A.} \bibnamefont{Renner}},
  \bibnamefont{and} \bibinfo{author}{\bibfnamefont{K.}~\bibnamefont{Sakurai}}
  (\bibinfo{year}{2015}), \eprint{arXiv:1512.07645}.

\bibitem[{\citenamefont{Davoudiasl and Zhang}(2015)}]{Davoudiasl:2015cuo}
\bibinfo{author}{\bibfnamefont{H.}~\bibnamefont{Davoudiasl}} \bibnamefont{and}
  \bibinfo{author}{\bibfnamefont{C.}~\bibnamefont{Zhang}}
  (\bibinfo{year}{2015}), \eprint{arXiv:1512.07672}.

\bibitem[{\citenamefont{Craig et~al.}(2015)\citenamefont{Craig, Draper, Kilic,
  and Thomas}}]{Craig:2015lra}
\bibinfo{author}{\bibfnamefont{N.}~\bibnamefont{Craig}},
  \bibinfo{author}{\bibfnamefont{P.}~\bibnamefont{Draper}},
  \bibinfo{author}{\bibfnamefont{C.}~\bibnamefont{Kilic}}, \bibnamefont{and}
  \bibinfo{author}{\bibfnamefont{S.}~\bibnamefont{Thomas}}
  (\bibinfo{year}{2015}), \eprint{arXiv:1512.07733}.

\bibitem[{\citenamefont{Das and Rai}(2015)}]{Das:2015enc}
\bibinfo{author}{\bibfnamefont{K.}~\bibnamefont{Das}} \bibnamefont{and}
  \bibinfo{author}{\bibfnamefont{S.~K.} \bibnamefont{Rai}}
  (\bibinfo{year}{2015}), \eprint{arXiv:1512.07789}.

\bibitem[{\citenamefont{Cheung et~al.}(2015)\citenamefont{Cheung, Ko, Lee,
  Park, and Tseng}}]{Cheung:2015cug}
\bibinfo{author}{\bibfnamefont{K.}~\bibnamefont{Cheung}},
  \bibinfo{author}{\bibfnamefont{P.}~\bibnamefont{Ko}},
  \bibinfo{author}{\bibfnamefont{J.~S.} \bibnamefont{Lee}},
  \bibinfo{author}{\bibfnamefont{J.}~\bibnamefont{Park}}, \bibnamefont{and}
  \bibinfo{author}{\bibfnamefont{P.-Y.} \bibnamefont{Tseng}}
  (\bibinfo{year}{2015}), \eprint{arXiv:1512.07853}.

\bibitem[{\citenamefont{Liu et~al.}(2015)\citenamefont{Liu, Wang, and
  Xue}}]{Liu:2015yec}
\bibinfo{author}{\bibfnamefont{J.}~\bibnamefont{Liu}},
  \bibinfo{author}{\bibfnamefont{X.-P.} \bibnamefont{Wang}}, \bibnamefont{and}
  \bibinfo{author}{\bibfnamefont{W.}~\bibnamefont{Xue}} (\bibinfo{year}{2015}),
  \eprint{arXiv:1512.07885}.

\bibitem[{\citenamefont{Zhang and Zhou}(2015)}]{Zhang:2015uuo}
\bibinfo{author}{\bibfnamefont{J.}~\bibnamefont{Zhang}} \bibnamefont{and}
  \bibinfo{author}{\bibfnamefont{S.}~\bibnamefont{Zhou}}
  (\bibinfo{year}{2015}), \eprint{arXiv:1512.07889}.

\bibitem[{\citenamefont{Casas et~al.}(2015)\citenamefont{Casas, Espinosa, and
  Moreno}}]{Casas:2015blx}
\bibinfo{author}{\bibfnamefont{J.~A.} \bibnamefont{Casas}},
  \bibinfo{author}{\bibfnamefont{J.~R.} \bibnamefont{Espinosa}},
  \bibnamefont{and} \bibinfo{author}{\bibfnamefont{J.~M.} \bibnamefont{Moreno}}
  (\bibinfo{year}{2015}), \eprint{arXiv:1512.07895}.

\bibitem[{\citenamefont{Hall et~al.}(2015)\citenamefont{Hall, Harigaya, and
  Nomura}}]{Hall:2015xds}
\bibinfo{author}{\bibfnamefont{L.~J.} \bibnamefont{Hall}},
  \bibinfo{author}{\bibfnamefont{K.}~\bibnamefont{Harigaya}}, \bibnamefont{and}
  \bibinfo{author}{\bibfnamefont{Y.}~\bibnamefont{Nomura}}
  (\bibinfo{year}{2015}), \eprint{arXiv:1512.07904}.

\bibitem[{\citenamefont{Gallicchio and Schwartz}(2011)}]{1106.3076}
\bibinfo{author}{\bibfnamefont{J.}~\bibnamefont{Gallicchio}} \bibnamefont{and}
  \bibinfo{author}{\bibfnamefont{M.~D.} \bibnamefont{Schwartz}},
  \bibinfo{journal}{Phys. Rev. Lett.} \textbf{\bibinfo{volume}{107}},
  \bibinfo{pages}{172001} (\bibinfo{year}{2011}), \eprint{arXiv:1106.3076}.

\bibitem[{\citenamefont{Larkoski et~al.}(2014)\citenamefont{Larkoski, Thaler,
  and Waalewijn}}]{1408.3122}
\bibinfo{author}{\bibfnamefont{A.~J.} \bibnamefont{Larkoski}},
  \bibinfo{author}{\bibfnamefont{J.}~\bibnamefont{Thaler}}, \bibnamefont{and}
  \bibinfo{author}{\bibfnamefont{W.~J.} \bibnamefont{Waalewijn}},
  \bibinfo{journal}{JHEP} \textbf{\bibinfo{volume}{11}}, \bibinfo{pages}{129}
  (\bibinfo{year}{2014}), \eprint{arXiv:1408.3122}.

\bibitem[{\citenamefont{Bhattacherjee et~al.}(2015)\citenamefont{Bhattacherjee,
  Mukhopadhyay, Nojiri, Sakaki, and Webber}}]{1501.04794}
\bibinfo{author}{\bibfnamefont{B.}~\bibnamefont{Bhattacherjee}},
  \bibinfo{author}{\bibfnamefont{S.}~\bibnamefont{Mukhopadhyay}},
  \bibinfo{author}{\bibfnamefont{M.~M.} \bibnamefont{Nojiri}},
  \bibinfo{author}{\bibfnamefont{Y.}~\bibnamefont{Sakaki}}, \bibnamefont{and}
  \bibinfo{author}{\bibfnamefont{B.~R.} \bibnamefont{Webber}},
  \bibinfo{journal}{JHEP} \textbf{\bibinfo{volume}{04}}, \bibinfo{pages}{131}
  (\bibinfo{year}{2015}), \eprint{arXiv:1501.04794}.

\bibitem[{\citenamefont{Goncalves et~al.}(2015)\citenamefont{Goncalves, Krauss,
  and Linten}}]{1512.05265}
\bibinfo{author}{\bibfnamefont{D.}~\bibnamefont{Goncalves}},
  \bibinfo{author}{\bibfnamefont{F.}~\bibnamefont{Krauss}}, \bibnamefont{and}
  \bibinfo{author}{\bibfnamefont{R.}~\bibnamefont{Linten}}
  (\bibinfo{year}{2015}), \eprint{arXiv:1512.05265}.

\bibitem[{\citenamefont{Sung}(2009)}]{0908.3688}
\bibinfo{author}{\bibfnamefont{I.}~\bibnamefont{Sung}}, \bibinfo{journal}{Phys.
  Rev.} \textbf{\bibinfo{volume}{D80}}, \bibinfo{pages}{094020}
  (\bibinfo{year}{2009}), \eprint{0908.3688}.

\bibitem[{\citenamefont{Ask et~al.}(2012)\citenamefont{Ask, Collins, Forshaw,
  Joshi, and Pilkington}}]{1108.2396}
\bibinfo{author}{\bibfnamefont{S.}~\bibnamefont{Ask}},
  \bibinfo{author}{\bibfnamefont{J.~H.} \bibnamefont{Collins}},
  \bibinfo{author}{\bibfnamefont{J.~R.} \bibnamefont{Forshaw}},
  \bibinfo{author}{\bibfnamefont{K.}~\bibnamefont{Joshi}}, \bibnamefont{and}
  \bibinfo{author}{\bibfnamefont{A.~D.} \bibnamefont{Pilkington}},
  \bibinfo{journal}{JHEP} \textbf{\bibinfo{volume}{01}}, \bibinfo{pages}{018}
  (\bibinfo{year}{2012}), \eprint{arXiv:1108.2396}.

\bibitem[{\citenamefont{Krohn et~al.}(2011)\citenamefont{Krohn, Randall, and
  Wang}}]{1101.0810}
\bibinfo{author}{\bibfnamefont{D.}~\bibnamefont{Krohn}},
  \bibinfo{author}{\bibfnamefont{L.}~\bibnamefont{Randall}}, \bibnamefont{and}
  \bibinfo{author}{\bibfnamefont{L.-T.} \bibnamefont{Wang}}
  (\bibinfo{year}{2011}), \eprint{arXiv:1101.0810}.

\bibitem[{\citenamefont{Butterworth et~al.}(2015)}]{Butterworth:2015oua}
\bibinfo{author}{\bibfnamefont{J.}~\bibnamefont{Butterworth}}
  \bibnamefont{et~al.} (\bibinfo{year}{2015}), \eprint{arXiv:1510.03865}.

\bibitem[{\citenamefont{Cieri et~al.}(2015)\citenamefont{Cieri, Coradeschi, and
  de~Florian}}]{1505.03162}
\bibinfo{author}{\bibfnamefont{L.}~\bibnamefont{Cieri}},
  \bibinfo{author}{\bibfnamefont{F.}~\bibnamefont{Coradeschi}},
  \bibnamefont{and}
  \bibinfo{author}{\bibfnamefont{D.}~\bibnamefont{de~Florian}},
  \bibinfo{journal}{JHEP} \textbf{\bibinfo{volume}{06}}, \bibinfo{pages}{185}
  (\bibinfo{year}{2015}), \eprint{arXiv:1505.03162}.

\bibitem[{\citenamefont{Catani et~al.}(2012)\citenamefont{Catani, Cieri,
  de~Florian, Ferrera, and Grazzini}}]{1110.2375}
\bibinfo{author}{\bibfnamefont{S.}~\bibnamefont{Catani}},
  \bibinfo{author}{\bibfnamefont{L.}~\bibnamefont{Cieri}},
  \bibinfo{author}{\bibfnamefont{D.}~\bibnamefont{de~Florian}},
  \bibinfo{author}{\bibfnamefont{G.}~\bibnamefont{Ferrera}}, \bibnamefont{and}
  \bibinfo{author}{\bibfnamefont{M.}~\bibnamefont{Grazzini}},
  \bibinfo{journal}{Phys. Rev. Lett.} \textbf{\bibinfo{volume}{108}},
  \bibinfo{pages}{072001} (\bibinfo{year}{2012}), \eprint{arXiv:1110.2375}.

\bibitem[{\citenamefont{Dokshitzer et~al.}(1978)\citenamefont{Dokshitzer,
  Diakonov, and Troian}}]{PHLTA.B79.269}
\bibinfo{author}{\bibfnamefont{Y.~L.} \bibnamefont{Dokshitzer}},
  \bibinfo{author}{\bibfnamefont{D.}~\bibnamefont{Diakonov}}, \bibnamefont{and}
  \bibinfo{author}{\bibfnamefont{S.~I.} \bibnamefont{Troian}},
  \bibinfo{journal}{Phys. Lett.} \textbf{\bibinfo{volume}{B79}},
  \bibinfo{pages}{269} (\bibinfo{year}{1978}).

\bibitem[{\citenamefont{Parisi and Petronzio}(1979)}]{NUPHA.B154.427}
\bibinfo{author}{\bibfnamefont{G.}~\bibnamefont{Parisi}} \bibnamefont{and}
  \bibinfo{author}{\bibfnamefont{R.}~\bibnamefont{Petronzio}},
  \bibinfo{journal}{Nucl. Phys.} \textbf{\bibinfo{volume}{B154}},
  \bibinfo{pages}{427} (\bibinfo{year}{1979}).

\bibitem[{\citenamefont{Collins and Soper}(1981)}]{NUPHA.B193.381}
\bibinfo{author}{\bibfnamefont{J.~C.} \bibnamefont{Collins}} \bibnamefont{and}
  \bibinfo{author}{\bibfnamefont{D.~E.} \bibnamefont{Soper}},
  \bibinfo{journal}{Nucl. Phys.} \textbf{\bibinfo{volume}{B193}},
  \bibinfo{pages}{381} (\bibinfo{year}{1981}), \bibinfo{note}{[Erratum: Nucl.
  Phys.B213,545(1983)]}.

\bibitem[{\citenamefont{Collins and Soper}(1982)}]{NUPHA.B197.446}
\bibinfo{author}{\bibfnamefont{J.~C.} \bibnamefont{Collins}} \bibnamefont{and}
  \bibinfo{author}{\bibfnamefont{D.~E.} \bibnamefont{Soper}},
  \bibinfo{journal}{Nucl. Phys.} \textbf{\bibinfo{volume}{B197}},
  \bibinfo{pages}{446} (\bibinfo{year}{1982}).

\bibitem[{\citenamefont{Collins et~al.}(1985)\citenamefont{Collins, Soper, and
  Sterman}}]{NUPHA.B250.199}
\bibinfo{author}{\bibfnamefont{J.~C.} \bibnamefont{Collins}},
  \bibinfo{author}{\bibfnamefont{D.~E.} \bibnamefont{Soper}}, \bibnamefont{and}
  \bibinfo{author}{\bibfnamefont{G.~F.} \bibnamefont{Sterman}},
  \bibinfo{journal}{Nucl. Phys.} \textbf{\bibinfo{volume}{B250}},
  \bibinfo{pages}{199} (\bibinfo{year}{1985}).

\bibitem[{\citenamefont{Catani et~al.}(2001)\citenamefont{Catani, de~Florian,
  and Grazzini}}]{hep-ph/0008184}
\bibinfo{author}{\bibfnamefont{S.}~\bibnamefont{Catani}},
  \bibinfo{author}{\bibfnamefont{D.}~\bibnamefont{de~Florian}},
  \bibnamefont{and} \bibinfo{author}{\bibfnamefont{M.}~\bibnamefont{Grazzini}},
  \bibinfo{journal}{Nucl. Phys.} \textbf{\bibinfo{volume}{B596}},
  \bibinfo{pages}{299} (\bibinfo{year}{2001}), \eprint{hep-ph/0008184}.

\bibitem[{\citenamefont{Kodaira and Trentadue}(1982)}]{PHLTA.B112.66}
\bibinfo{author}{\bibfnamefont{J.}~\bibnamefont{Kodaira}} \bibnamefont{and}
  \bibinfo{author}{\bibfnamefont{L.}~\bibnamefont{Trentadue}},
  \bibinfo{journal}{Phys. Lett.} \textbf{\bibinfo{volume}{B112}},
  \bibinfo{pages}{66} (\bibinfo{year}{1982}).

\bibitem[{\citenamefont{Kodaira and Trentadue}(1983)}]{PHLTA.B123.335}
\bibinfo{author}{\bibfnamefont{J.}~\bibnamefont{Kodaira}} \bibnamefont{and}
  \bibinfo{author}{\bibfnamefont{L.}~\bibnamefont{Trentadue}},
  \bibinfo{journal}{Phys. Lett.} \textbf{\bibinfo{volume}{B123}},
  \bibinfo{pages}{335} (\bibinfo{year}{1983}).

\bibitem[{\citenamefont{Davies and Stirling}(1984)}]{NUPHA.B244.337}
\bibinfo{author}{\bibfnamefont{C.~T.~H.} \bibnamefont{Davies}}
  \bibnamefont{and} \bibinfo{author}{\bibfnamefont{W.~J.}
  \bibnamefont{Stirling}}, \bibinfo{journal}{Nucl. Phys.}
  \textbf{\bibinfo{volume}{B244}}, \bibinfo{pages}{337} (\bibinfo{year}{1984}).

\bibitem[{\citenamefont{Ladinsky and Yuan}(1994)}]{hep-ph/9311341}
\bibinfo{author}{\bibfnamefont{G.~A.} \bibnamefont{Ladinsky}} \bibnamefont{and}
  \bibinfo{author}{\bibfnamefont{C.~P.} \bibnamefont{Yuan}},
  \bibinfo{journal}{Phys. Rev.} \textbf{\bibinfo{volume}{D50}},
  \bibinfo{pages}{4239} (\bibinfo{year}{1994}), \eprint{hep-ph/9311341}.

\bibitem[{\citenamefont{Qiu and Zhang}(2001)}]{hep-ph/0012348}
\bibinfo{author}{\bibfnamefont{J.-w.} \bibnamefont{Qiu}} \bibnamefont{and}
  \bibinfo{author}{\bibfnamefont{X.-f.} \bibnamefont{Zhang}},
  \bibinfo{journal}{Phys. Rev.} \textbf{\bibinfo{volume}{D63}},
  \bibinfo{pages}{114011} (\bibinfo{year}{2001}), \eprint{hep-ph/0012348}.

\bibitem[{\citenamefont{Bozzi et~al.}(2006)\citenamefont{Bozzi, Catani,
  de~Florian, and Grazzini}}]{hep-ph/0508068}
\bibinfo{author}{\bibfnamefont{G.}~\bibnamefont{Bozzi}},
  \bibinfo{author}{\bibfnamefont{S.}~\bibnamefont{Catani}},
  \bibinfo{author}{\bibfnamefont{D.}~\bibnamefont{de~Florian}},
  \bibnamefont{and} \bibinfo{author}{\bibfnamefont{M.}~\bibnamefont{Grazzini}},
  \bibinfo{journal}{Nucl. Phys.} \textbf{\bibinfo{volume}{B737}},
  \bibinfo{pages}{73} (\bibinfo{year}{2006}), \eprint{hep-ph/0508068}.

\bibitem[{\citenamefont{Bauer et~al.}(2000)\citenamefont{Bauer, Fleming, and
  Luke}}]{hep-ph/0005275}
\bibinfo{author}{\bibfnamefont{C.~W.} \bibnamefont{Bauer}},
  \bibinfo{author}{\bibfnamefont{S.}~\bibnamefont{Fleming}}, \bibnamefont{and}
  \bibinfo{author}{\bibfnamefont{M.~E.} \bibnamefont{Luke}},
  \bibinfo{journal}{Phys. Rev.} \textbf{\bibinfo{volume}{D63}},
  \bibinfo{pages}{014006} (\bibinfo{year}{2000}), \eprint{hep-ph/0005275}.

\bibitem[{\citenamefont{Bauer et~al.}(2001)\citenamefont{Bauer, Fleming,
  Pirjol, and Stewart}}]{hep-ph/0011336}
\bibinfo{author}{\bibfnamefont{C.~W.} \bibnamefont{Bauer}},
  \bibinfo{author}{\bibfnamefont{S.}~\bibnamefont{Fleming}},
  \bibinfo{author}{\bibfnamefont{D.}~\bibnamefont{Pirjol}}, \bibnamefont{and}
  \bibinfo{author}{\bibfnamefont{I.~W.} \bibnamefont{Stewart}},
  \bibinfo{journal}{Phys. Rev.} \textbf{\bibinfo{volume}{D63}},
  \bibinfo{pages}{114020} (\bibinfo{year}{2001}), \eprint{hep-ph/0011336}.

\bibitem[{\citenamefont{Bauer et~al.}(2002{\natexlab{a}})\citenamefont{Bauer,
  Pirjol, and Stewart}}]{hep-ph/0109045}
\bibinfo{author}{\bibfnamefont{C.~W.} \bibnamefont{Bauer}},
  \bibinfo{author}{\bibfnamefont{D.}~\bibnamefont{Pirjol}}, \bibnamefont{and}
  \bibinfo{author}{\bibfnamefont{I.~W.} \bibnamefont{Stewart}},
  \bibinfo{journal}{Phys. Rev.} \textbf{\bibinfo{volume}{D65}},
  \bibinfo{pages}{054022} (\bibinfo{year}{2002}{\natexlab{a}}),
  \eprint{hep-ph/0109045}.

\bibitem[{\citenamefont{Bauer et~al.}(2002{\natexlab{b}})\citenamefont{Bauer,
  Fleming, Pirjol, Rothstein, and Stewart}}]{hep-ph/0202088}
\bibinfo{author}{\bibfnamefont{C.~W.} \bibnamefont{Bauer}},
  \bibinfo{author}{\bibfnamefont{S.}~\bibnamefont{Fleming}},
  \bibinfo{author}{\bibfnamefont{D.}~\bibnamefont{Pirjol}},
  \bibinfo{author}{\bibfnamefont{I.~Z.} \bibnamefont{Rothstein}},
  \bibnamefont{and} \bibinfo{author}{\bibfnamefont{I.~W.}
  \bibnamefont{Stewart}}, \bibinfo{journal}{Phys. Rev.}
  \textbf{\bibinfo{volume}{D66}}, \bibinfo{pages}{014017}
  (\bibinfo{year}{2002}{\natexlab{b}}), \eprint{hep-ph/0202088}.

\bibitem[{\citenamefont{Bozzi et~al.}(2003)\citenamefont{Bozzi, Catani,
  de~Florian, and Grazzini}}]{hep-ph/0302104}
\bibinfo{author}{\bibfnamefont{G.}~\bibnamefont{Bozzi}},
  \bibinfo{author}{\bibfnamefont{S.}~\bibnamefont{Catani}},
  \bibinfo{author}{\bibfnamefont{D.}~\bibnamefont{de~Florian}},
  \bibnamefont{and} \bibinfo{author}{\bibfnamefont{M.}~\bibnamefont{Grazzini}},
  \bibinfo{journal}{Phys. Lett.} \textbf{\bibinfo{volume}{B564}},
  \bibinfo{pages}{65} (\bibinfo{year}{2003}), \eprint{hep-ph/0302104}.

\bibitem[{\citenamefont{de~Florian et~al.}(2011)\citenamefont{de~Florian,
  Ferrera, Grazzini, and Tommasini}}]{1109.2109}
\bibinfo{author}{\bibfnamefont{D.}~\bibnamefont{de~Florian}},
  \bibinfo{author}{\bibfnamefont{G.}~\bibnamefont{Ferrera}},
  \bibinfo{author}{\bibfnamefont{M.}~\bibnamefont{Grazzini}}, \bibnamefont{and}
  \bibinfo{author}{\bibfnamefont{D.}~\bibnamefont{Tommasini}},
  \bibinfo{journal}{JHEP} \textbf{\bibinfo{volume}{11}}, \bibinfo{pages}{064}
  (\bibinfo{year}{2011}), \eprint{arXiv:1109.2109}.

\bibitem[{\citenamefont{Bozzi et~al.}(2011)\citenamefont{Bozzi, Catani,
  Ferrera, de~Florian, and Grazzini}}]{1007.2351}
\bibinfo{author}{\bibfnamefont{G.}~\bibnamefont{Bozzi}},
  \bibinfo{author}{\bibfnamefont{S.}~\bibnamefont{Catani}},
  \bibinfo{author}{\bibfnamefont{G.}~\bibnamefont{Ferrera}},
  \bibinfo{author}{\bibfnamefont{D.}~\bibnamefont{de~Florian}},
  \bibnamefont{and} \bibinfo{author}{\bibfnamefont{M.}~\bibnamefont{Grazzini}},
  \bibinfo{journal}{Phys. Lett.} \textbf{\bibinfo{volume}{B696}},
  \bibinfo{pages}{207} (\bibinfo{year}{2011}), \eprint{arXiv:1007.2351}.

\bibitem[{\citenamefont{Becher and Neubert}(2011)}]{1007.4005}
\bibinfo{author}{\bibfnamefont{T.}~\bibnamefont{Becher}} \bibnamefont{and}
  \bibinfo{author}{\bibfnamefont{M.}~\bibnamefont{Neubert}},
  \bibinfo{journal}{Eur. Phys. J.} \textbf{\bibinfo{volume}{C71}},
  \bibinfo{pages}{1665} (\bibinfo{year}{2011}), \eprint{arXiv:1007.4005}.

\bibitem[{\citenamefont{Becher et~al.}(2013)\citenamefont{Becher, Neubert, and
  Wilhelm}}]{1212.2621}
\bibinfo{author}{\bibfnamefont{T.}~\bibnamefont{Becher}},
  \bibinfo{author}{\bibfnamefont{M.}~\bibnamefont{Neubert}}, \bibnamefont{and}
  \bibinfo{author}{\bibfnamefont{D.}~\bibnamefont{Wilhelm}},
  \bibinfo{journal}{JHEP} \textbf{\bibinfo{volume}{05}}, \bibinfo{pages}{110}
  (\bibinfo{year}{2013}), \eprint{arXiv:1212.2621}.

\bibitem[{\citenamefont{Alwall et~al.}(2014)\citenamefont{Alwall, Frederix,
  Frixione, Hirschi, Maltoni et~al.}}]{Alwall:2014hca}
\bibinfo{author}{\bibfnamefont{J.}~\bibnamefont{Alwall}},
  \bibinfo{author}{\bibfnamefont{R.}~\bibnamefont{Frederix}},
  \bibinfo{author}{\bibfnamefont{S.}~\bibnamefont{Frixione}},
  \bibinfo{author}{\bibfnamefont{V.}~\bibnamefont{Hirschi}},
  \bibinfo{author}{\bibfnamefont{F.}~\bibnamefont{Maltoni}},
  \bibnamefont{et~al.}, \bibinfo{journal}{JHEP}
  \textbf{\bibinfo{volume}{1407}}, \bibinfo{pages}{079} (\bibinfo{year}{2014}),
  \eprint{arXiv:1405.0301}.

\bibitem[{\citenamefont{Dulat et~al.}(2015)\citenamefont{Dulat, Hou, Gao,
  Guzzi, Huston, Nadolsky, Pumplin, Schmidt, Stump, and Yuan}}]{Dulat:2015mca}
\bibinfo{author}{\bibfnamefont{S.}~\bibnamefont{Dulat}},
  \bibinfo{author}{\bibfnamefont{T.~J.} \bibnamefont{Hou}},
  \bibinfo{author}{\bibfnamefont{J.}~\bibnamefont{Gao}},
  \bibinfo{author}{\bibfnamefont{M.}~\bibnamefont{Guzzi}},
  \bibinfo{author}{\bibfnamefont{J.}~\bibnamefont{Huston}},
  \bibinfo{author}{\bibfnamefont{P.}~\bibnamefont{Nadolsky}},
  \bibinfo{author}{\bibfnamefont{J.}~\bibnamefont{Pumplin}},
  \bibinfo{author}{\bibfnamefont{C.}~\bibnamefont{Schmidt}},
  \bibinfo{author}{\bibfnamefont{D.}~\bibnamefont{Stump}}, \bibnamefont{and}
  \bibinfo{author}{\bibfnamefont{C.~P.} \bibnamefont{Yuan}}
  (\bibinfo{year}{2015}), \eprint{arXiv:1506.07443}.

\bibitem[{\citenamefont{Sjostrand et~al.}(2006)\citenamefont{Sjostrand, Mrenna,
  and Skands}}]{Sjostrand:2006za}
\bibinfo{author}{\bibfnamefont{T.}~\bibnamefont{Sjostrand}},
  \bibinfo{author}{\bibfnamefont{S.}~\bibnamefont{Mrenna}}, \bibnamefont{and}
  \bibinfo{author}{\bibfnamefont{P.~Z.} \bibnamefont{Skands}},
  \bibinfo{journal}{JHEP} \textbf{\bibinfo{volume}{0605}}, \bibinfo{pages}{026}
  (\bibinfo{year}{2006}), \eprint{hep-ph/0603175}.

\bibitem[{\citenamefont{Field}(2011)}]{Field:2011iq}
\bibinfo{author}{\bibfnamefont{R.}~\bibnamefont{Field}}, \bibinfo{journal}{Acta
  Phys.Polon.} \textbf{\bibinfo{volume}{B42}}, \bibinfo{pages}{2631}
  (\bibinfo{year}{2011}), \eprint{arXiv:1110.5530}.

\bibitem[{\citenamefont{de~Favereau et~al.}(2014)}]{deFavereau:2013fsa}
\bibinfo{author}{\bibfnamefont{J.}~\bibnamefont{de~Favereau}}
  \bibnamefont{et~al.} (\bibinfo{collaboration}{DELPHES 3}),
  \bibinfo{journal}{JHEP} \textbf{\bibinfo{volume}{1402}}, \bibinfo{pages}{057}
  (\bibinfo{year}{2014}), \eprint{arXiv:1307.6346}.

\bibitem[{\citenamefont{Cacciari et~al.}(2012)\citenamefont{Cacciari, Salam,
  and Soyez}}]{Cacciari:2011ma}
\bibinfo{author}{\bibfnamefont{M.}~\bibnamefont{Cacciari}},
  \bibinfo{author}{\bibfnamefont{G.~P.} \bibnamefont{Salam}}, \bibnamefont{and}
  \bibinfo{author}{\bibfnamefont{G.}~\bibnamefont{Soyez}},
  \bibinfo{journal}{Eur.Phys.J.} \textbf{\bibinfo{volume}{C72}},
  \bibinfo{pages}{1896} (\bibinfo{year}{2012}), \eprint{arXiv:1111.6097}.

\bibitem[{ATL(2015{\natexlab{b}})}]{ATL-PHYS-PUB-2015-022}
\bibinfo{type}{Tech. Rep.} \bibinfo{number}{ATL-PHYS-PUB-2015-022},
  \bibinfo{institution}{CERN}, \bibinfo{address}{Geneva}
  (\bibinfo{year}{2015}{\natexlab{b}}),
  \urlprefix\url{http://cds.cern.ch/record/2037697}.

\bibitem[{\citenamefont{Read}(2002)}]{Read:2002hq}
\bibinfo{author}{\bibfnamefont{A.~L.} \bibnamefont{Read}}, \bibinfo{journal}{J.
  Phys.} \textbf{\bibinfo{volume}{G28}}, \bibinfo{pages}{2693}
  (\bibinfo{year}{2002}).

\bibitem[{\citenamefont{Cowan et~al.}(2011)\citenamefont{Cowan, Cranmer, Gross,
  and Vitells}}]{Cowan:2010js}
\bibinfo{author}{\bibfnamefont{G.}~\bibnamefont{Cowan}},
  \bibinfo{author}{\bibfnamefont{K.}~\bibnamefont{Cranmer}},
  \bibinfo{author}{\bibfnamefont{E.}~\bibnamefont{Gross}}, \bibnamefont{and}
  \bibinfo{author}{\bibfnamefont{O.}~\bibnamefont{Vitells}},
  \bibinfo{journal}{Eur.Phys.J.} \textbf{\bibinfo{volume}{C71}},
  \bibinfo{pages}{1554} (\bibinfo{year}{2011}), \eprint{arXiv:1007.1727}.

\end{thebibliography}

\end{document}